\documentclass{emulateapj}
\usepackage{graphics}

\usepackage[usenames]{color}

\slugcomment{Last revision \today}

\newcommand{\Ntwo}{$N_{\rm 200}$}
\newcommand{\Mtwo}{$M_{\rm 200}$}

\newcommand{\rtwo}{$r_{200}$}
\newcommand{\rad}{$r/r_{200}$}
\newcommand{\Ltwo}{$L_{200}$}
\newcommand{\Lbcg}{$^{0.25}L_{BCG}$}
\newcommand{\lbcg}{$^{0.25}L_{BCG}$}
\newcommand{\Lstar}{$L_*$}
\newcommand{\lstar}{$L_*$}
\newcommand{\Lstarsat}{$^{0.25}L_*(sat)$}

\newcommand{\gmr}{$^{0.25}(g-r)$}

\newcommand{\fred}{$f_{\rm R}$}
\newcommand{\Mi}{$^{0.25}M_i$ - 5log$_{10} h$}
\newcommand{\Msun}{$h^{-1} M_{\odot}$}

\newcommand{\Mpc}{$h^{-1}$ Mpc}
\newcommand{\ie}{{\em i.e.}}
\newcommand{\eg}{{\em e.g.}}
\newcommand{\nclust}{165,597}

\newcommand{\nvir}{$N_{200}$}

\newcommand{\maxbcg}{MaxBCG}

\newcommand{\ugriz}{$u, g, r, i, z$}

\newcommand{\altnumpairsTenMpc}{2.7 billion}

\newcommand{\gae}{\lower 2pt \hbox{$\, \buildrel {\scriptstyle >}\over {\scriptstyle
\sim}\,$}}
\newcommand{\lae}{\lower 2pt \hbox{$\, \buildrel {\scriptstyle <}\over {\scriptstyle
\sim}\,$}}

\shorttitle{The Galaxy Content of SDSS Clusters and Groups}
\shortauthors{Hansen et al.}

\begin{document}

\title{The Galaxy Content of SDSS Clusters and Groups}

\author{Sarah M. Hansen\altaffilmark{1,2},
Erin S. Sheldon\altaffilmark{3},
Risa H. Wechsler\altaffilmark{4}, and
Benjamin P. Koester\altaffilmark{1}
}

\altaffiltext{1}{Department of Astronomy and Astrophysics,
University of Chicago, Chicago, IL 60637; {\tt shansen@kicp.uchicago.edu}}
\altaffiltext{2}{Kavli Institute for Cosmological Physics,
University of Chicago, Chicago, IL 60637}
\altaffiltext{3}{Center for Cosmology and Particle Physics, Department of Physics, New York University, 4 Washington Place, New York, NY 10003}
\altaffiltext{4}{Kavli Institute for Particle Astrophysics \& Cosmology,
  Physics Department, and Stanford Linear Accelerator Center,
  Stanford University, Stanford, CA 94305}

\begin{abstract}

Imaging data from the Sloan Digital Sky Survey are used to
characterize the population of galaxies in groups and clusters
detected with the MaxBCG algorithm. We investigate the dependence of
Brightest Cluster Galaxy (BCG) luminosity, and the distributions of
satellite galaxy luminosity and satellite color, on cluster properties
over the redshift range $0.1 \le z \le 0.3$. The size of the dataset
allows us to make measurements in many bins of cluster richness,
radius and redshift. We find that, within \rtwo\ of clusters with
mass above $3 \times 10^{13}$\Msun, the luminosity function of both
red and blue satellites is only weakly dependent on richness. We
further find that the shape of the satellite luminosity function does
not depend on cluster-centric distance for magnitudes brighter than
\Mi\ $= -19$. However, the mix of faint red and blue galaxies
changes dramatically. The satellite red fraction is dependent on
cluster-centric distance, galaxy luminosity and cluster mass, and
also increases by $\sim$5\% between redshifts 0.28 and 0.2, independent of
richness. We find that BCG luminosity is tightly correlated with
cluster richness, scaling as $L_{BCG} \sim M_{200}^{0.3}$, and has a Gaussian
distribution at fixed richness, with $\sigma_{log L} \sim 0.17$ for massive
clusters.  The ratios of BCG luminosity to total cluster luminosity
and characteristic satellite luminosity scale strongly with cluster
richness: in richer systems, BCGs contribute a smaller fraction of the
total light, but are brighter compared to typical satellites. This
study demonstrates the power of cross-correlation techniques for
measuring galaxy populations in purely photometric data.

\end{abstract}


\section{Introduction}\label{sec:intro} 

Clusters of galaxies are important systems for studying both galaxy evolution
and cosmology. Used as laboratories with well-defined environments, these
massive objects are a tool for investigating processes that influence galaxies'
physical characteristics. Used as tracers of the underlying mass distribution,
they are a tool for investigating the evolution of structure and the nature of
dark energy in the Universe.  These objectives are closely linked: cosmological
studies require accurate knowledge of cluster selection and redshift- and
mass-observable relations, and these facts are directly related the
evolutionary properties of the cluster galaxy population.

In the context of galaxy evolution, the high-density environment of
galaxy clusters is a particularly interesting place to examine the
galaxy population.  Several studies have suggested substantial galaxy
transformation in such environments.  Galaxy morphology,
star-formation rate, and luminosity have long been known to depend on
cluster properties and to depart significantly from the cosmological
average \citep[\eg,][]{Hubble26, Abell62, Oemler74, Dressler80}.
Historically, the cluster galaxy content has been quantified by the
luminosity function (LF) and type fraction (such as the late-type or
blue fraction).  Measurement of these quantities as a function of
cluster mass, redshift, and distance from the cluster center provides
insight into the underlying physical mechanisms responsible for these
trends.

While the LF is primarily a decreasing function of luminosity,
galaxies are bimodal in color and spectral type, with red, early-type
galaxies displaying little ongoing star formation, and blue, late-type
galaxies exhibiting signs of recent star formation
\citep[\eg,][]{Strateva01, Baldry04, Bell04, Menanteau06,
Blanton05b}.  This bimodality was in place by $z\sim1$ at the latest and
may provide a signpost of galaxy transformation
\citep[\eg,][]{Faber07}.  Although this bimodality persists in all
environments, the fraction of galaxies in each class
(a.k.a. the red and blue fractions) changes systematically with local
density; this trend is the so-called morphology-density relationship
\citep{Oemler74,Dressler80,Dressler97,Smith05}.  In recent large
galaxy surveys this work has been extended to show that a wide range
of galaxy properties, including morphology, star-formation rate, and
color, depend on local density
\citep{Gomez03,Balogh04a,Balogh04b,Hogg04,Kauffmann04,Tovmassian04,Blanton05b,Christlein05,Croton05,Rojas05,Cooper06,Mandelbaum06,Cooper07}.

The advent of large galaxy surveys has made it possible to place
observational constraints on both the type fraction and the LF as a
function of cluster mass (the conditional luminosity function, CLF),
and to further investigate how these quantities depend on other
variables.  In the Sloan Digital Sky Survey \citep*[][SDSS]{York00},
\citet{Goto02} and \citet{Hansen05} examined the LF as a function of
cluster richness and cluster-centric distance for systems found in the
SDSS Early Data Release, and \citet{Weinmann06b} measured the CLF measured
from a group catalog derived from the spectroscopic sample of SDSS
DR2.  Using the 2dF Galaxy Redshift Survey, \citet{depropris03} and
\citet{Robotham06} compared the LF in high- and low-mass systems; a
similar study was performed with a sample of 93 X-ray selected
clusters \citep{LMS04}.  The dependence of the LF on galaxy color has
been recently investigated \citep{PopessoLF} as has the galaxy type
fraction \citep{Goto03,depropris04} for clusters in these large surveys.

The type fraction of galaxies in clusters depends on
both cluster richness and redshift: the fraction of star-forming
galaxies at fixed local density is larger at higher redshift, an
effect known as the Butcher--Oemler effect
\citep{BO78,ButcherOemler}. This effect is now well-documented, if not
entirely well-explained, in clusters over a wide range of masses by
studies of the blue fraction, or its converse, the red fraction
\citep{Rakos95,Margoniner00,Ellingson01,KodamaBower01,Margoniner01,depropris04,Martinez06,Gerke07}.
Other indicators of galaxy state, including galaxy morphology and
emission line strength, also show a trend with redshift
\citep{AS93,ODB97,Balogh97,Couch98,vanDokkum00,Fasano00,Lubin02,Goto03,Treu03,Wilman05,Poggianti06,Desai07,vanderwel07}.
However, few samples to date have had both large numbers of systems
and well-understood mass proxies, so the dynamical range and mass
resolution of these previous works has been somewhat limited.

Another characteristic of galaxy clusters is the presence of a
highly-luminous galaxy near the cluster center --- the Brightest
Cluster Galaxy (BCG).  In addition to being extraordinarily luminous,
BCGs differ in a number of ways from other cluster members: they tend
to have extended light profiles \citep{Matthews64, Tonry87,
  Schombert88, Gonzalez00, Gonzalez03}, larger size at fixed
luminosity than other early types \citep[][and references
therein]{Bernardi07} and may contain a larger fraction of dark matter
than typical galaxies \citep[\eg,][]{Mandelbaum06,Anja07}.  Also,
while the traditional fitting function to the LF of
\citet{Schechter76} provides a good fit for satellite galaxies, BCGs
follow a different distribution, causing the so-called ``bright end
bump'' \citep[\eg,][]{Hansen05}. This difference between the BCG and
other satellites in a cluster is also manifested in the luminosity gap
statistic, the difference in luminosity between the BCG and the next
brightest cluster member, which may be indicative of the special
accretion history of BCGs \citep{Ostriker75, Tremaine77, Loh06}.
BCGs are also distinct from other galaxies with 
similar mass that are not at the center of cluster-sized potential
wells \citep{Anja07}. Indeed, the properties of BCGs seem to be
closely linked to properties of their host clusters
\citep{Sandage73,Sch83,Schombert88, Edge91, Brough02,
  Degrandi04,Brough05,Loh06}, including to the masses of their parent
halos \citep{LinMohr04, Mandelbaum06, ZCZ07}.  The outer light profile
of BCGs often merges smoothly with the diffuse intra-cluster light
(ICL), suggesting again a coupling between the BCG and the host halo
\citep{Gonzalez05}.  Generally the BCG+ICL light is closely linked
with cluster mass
\citep{Zibetti05,Conroy07ICL,Purcell07,Gonzalez07}. As BCGs have
properties different from those of the rest of the cluster members,
BCGs and satellites are often analyzed separately, and we follow this
convention here.

It is expected that the properties of cluster galaxies are closely
tied to the merging and accretion history of their parent dark matter
halos.  Current models based on Cold Dark Matter (CDM) suggest that
the stars in BCGs were formed in dense peaks quite early but that the
BCGs were assembled in a series of galaxy merging events that continue
until relatively recent times
\citep[e.g.][]{AS98,Dubinski98,Gao04a,BK06,DeLucia07}.  Satellite
galaxies in clusters are now generally understood to be hosted by
smaller dark matter halos that have merged into the parent halo.
Several features of the observed cluster galaxy populations can be
understood based on the assembly history of the parent halo
\citep[\eg,][]{Poggianti06, Iro07}.  One such characterization is the core distinction
between central galaxies and satellites: the central concentrations of
mass and light in massive halos continue to build up while the growth
of satellite systems is halted upon accretion.  BCG luminosity, and
the correlations between this luminosity and both cluster mass and
satellite properties, can thus provide insight into the assembly
histories of clusters.

The merger history of dark matter halos alone is not enough to
account for the bimodality in galaxy properties and its dependence on
redshift and environment.  Several physical processes have been
proposed to transform star-forming galaxies into the typical cluster
galaxies on the red sequence.  Although the relative strengths of
these processes are still hotly debated, a consensus is emerging that
the main transformation mechanisms are related to the mass of the
host halo and whether (and for how long) the galaxy has been a satellite
within a larger system.  Among the suggested transformation
mechanisms, some are expected to be most effective in rich clusters,
such as ram-pressure stripping \citep{GunnGott72}, interaction with
the cluster potential \citep{ByrdValtonen90} and high-velocity close
encounters \citep[``harassment;''][]{Moore96}. However, studies of
very poor systems have shown that the environmental dependence of
galaxy properties is not limited to the richest objects
\citep[\eg,][]{Zabludoff98,Weinmann06a,Gerke07}. Processes that can
operate efficiently in low velocity dispersion systems therefore must
also play a role in shaping the galaxy population: \eg, galaxy mergers
\citep{TT72} and ``strangulation,'' a cutoff to gas accretion onto
galaxy disks by stripping or AGN feedback
\citep{Larson80,BNM00,Croton06}. It is likely that some combination of
these effects is at work.  Distinguishing their relative significance
requires precisely quantifying cluster galaxy properties over a
wide range of masses and as a function of cluster-centric distance.  
These data will provide information on both the assembly histories of 
clusters and on the physical mechanisms that trigger and quench star 
formation.

Models of galaxy evolution in a cosmological context most readily
predict galaxy properties as a function of halo mass rather than
cluster observables.  In order to make these comparisons, a reliable
mass--observable relationship is a prerequisite. In addition, there is
consensus that the luminosity function and type fraction both depend
on a number of variables, complicating detailed comparison between
clusters of different mass.  For example, the cluster galaxy LF
depends on cluster-centric distance, and the size of the bound regions
of clusters scales with mass.  In order to make physically meaningful
comparisons between LFs of different mass clusters, an aperture scaled
to the bound region is therefore preferable to a fixed metric
aperture.  With recent extensive surveys providing well-calibrated
cluster catalogs spanning a wide range in mass, it is possible to
examine in detail the dependence of the cluster galaxy population on
several cluster and galaxy properties simultaneously.

Large, homogeneous photometric surveys such as the SDSS provide rich
data with which to characterize the cluster galaxy population.  These
data have been used to define large, robust, clean samples of galaxy
clusters with accurate photometric redshifts.  These samples are
sizable enough to split on several variables allowing detailed
statistical exploration of the galaxy populations in clusters.
Currently, the largest sample of clusters available is the \maxbcg\
catalog from the Sloan Digital Sky Survey \citep{Koester07a}.  The
selection effects of the cluster-finding algorithm are well understood
\citep{Koester07b, Rozo07a}, and there are a number of studies
exploring the mass--richness relationship for these objects
\citep{Rozo07b,Becker07,Sheldon07a, Johnston07b,Rykoff07}.  In
this work, we convert cluster richness to cluster mass using weak
lensing measurements of the \maxbcg\ mass--richness relation
\citep{Sheldon07a,Johnston07b}.  The quality and quantity of these
data allow for detailed measurements of a variety of cluster galaxy
properties as a function of cluster mass and radius.  For satellite
galaxies we then measure the luminosity function of all, red, and blue
satellites conditional on both mass and cluster radius, and
investigate the dependence of the red fraction of satellites on
cluster mass, redshift, galaxy luminosity and distance from cluster
center; for BCGs we quantify the dependence on cluster mass of both
the BCG luminosity and the relationship between the BCG luminosity and
satellite galaxy luminosities.  Although this sample of clusters
extends only to $z = 0.3$, these objects provide a valuable
low-redshift baseline with which higher redshift samples may be
compared.

In this paper we use a statistical background-subtraction technique to
measure mean galaxy properties over a wide range of color and
luminosity in \maxbcg\ clusters.  We average the signal from many
clusters binned by cluster properties, and statistically subtract the
contribution from random galaxies along the line of sight.  This
method of cross-correlating clusters with the galaxy population
provides very precise statistical measurements, and allows us to study
blue and low luminosity galaxies that are indistinguishable from the
background in individual clusters.  We test the background-correction
algorithm by running the full analysis on realistic mock catalogs, and
find that we are able to robustly recover 3D cluster properties using
these methods.  The statistical techniques presented here for
background correction of photometric data will be directly applicable
to future large multi-band imaging programs, including the Dark Energy
Survey \citep[DES\footnote{\tt
  http://www.darkenergysurvey.org},][]{DES} and the Large Synoptic
Survey Telescope \citep[LSST\footnote{{\tt
    http://www.lsst.org}},][]{LSST} and the Panoramic Survey Telescope
\& Rapid Response System \citep[Pan-STARRS\footnote{{\tt
    http://pan-starrs.ifa.hawaii.edu}},][]{PANSTARRS} and will be
essential for leveraging these data to provide the desired insights
into cosmology and galaxy evolution.

The paper is organized as follows: in \S\ \ref{sec:data} we describe
the SDSS and simulation data used; we present the stacking and
background-correction method in \S\ \ref{sec:methods}. Our primary
results are given in \S\ \ref{sec:results}: \S\ \ref{sec:sats} presents
the luminosity and color characteristics of the satellite population,
while \S\ \ref{sec:BCGs} discusses the BCG population. A summary and
discussion of the implications of the results is given in \S\
\ref{sec:conclusion}.

The notation used for cluster-related variables in previous
\maxbcg\ work includes defining \Ntwo\ and $L_{200}$ as the counts and
$i$-band luminosity of red-sequence galaxies within the measurement
aperture of the cluster finder and with L $> 0.4L_*$. We note that as
the aperture for cluster finding was determined with a previous
definition of richness, it is not strictly the true value of \rtwo\
for these systems (in fact, it is larger), and only red galaxies are
included in these definitions. In this work we will refer to the total
excess luminosity associated with the light from galaxies of {\em all}
colors above a luminosity threshold and within the measured \rtwo\ of
these systems as $L_{200}$.  In addition, we follow the standard convention of
using $R$ to to denote projected, 2D radii and $r$ to refer to
deprojected, 3D radii. Where necessary for computing distances, we
assume a flat, LCDM cosmology with H$_{0} = 100h$ km s$^{-1}$
Mpc$^{-1}$, $h = 0.7$, and matter density $\Omega_m = 0.3$.

\section{Data} \label{sec:data}

We use the photometric data of the Sloan Digital Sky Survey
(SDSS) Data Release 4. The SDSS
data were obtained using a specially-designed 2.5-m telescope \citep{Gunn06},
operated in drift-scan mode in five bandpasses (\ugriz) to a limiting magnitude
of $r<22.5$ \citep{Fukugita96, Gunn98, Lupton10, Hogg01, Smith02}. The survey
covers much of the North Galactic Cap and a small, repeatedly-scanned region in
the South. Apparent magnitudes are corrected for Galactic extinction using the
maps of \citet{Schlegel98}. Photometric errors at bright magnitudes are $\le
3\%$, limited by systematic uncertainty \citep{Ivezic04}; astrometric errors
are typically smaller than 50 mas per coordinate \citep{Pier03}. Further
details of the SDSS data may be found in the most recent data release paper of
\citet{dr6}.

With these data, we define a catalog of galaxies, then run our
cluster-finding algorithm on this list to produce a catalog of galaxy
clusters. The set of clusters and galaxies are then jointly analyzed
to determine the binned correlation function measurements used in this
work to describe the galaxy population associated with clusters. In this
section, we describe the galaxy and cluster catalogs.

\subsection{Galaxy Sample} 

Our photometric galaxy catalog was generated by applying the Bayesian
star-galaxy separation method developed in \citet{Scranton02} to the full set
of SDSS data. The primary source of confusion in star-galaxy separation at
faint magnitudes is shot noise, which causes stars to scatter out of the
stellar locus and galaxies to scatter into the stellar locus.  The
\citet{Scranton02} method uses knowledge about the underlying size distribution
of objects as a function of apparent magnitude and seeing to assign a
probability of being a galaxy to each object. We examined objects from the
repeatedly scanned SDSS southern stripe and co-added the fluxes from different
exposures at the catalog level to provide more precise sizes and magnitudes.
Using regions with at least 20 scans, we selected objects on single exposures
with typical observing conditions, but characterized their size distribution as
a function of magnitude and seeing with the co-add. From these distributions we
generated a galaxy probability for every object in the survey.  The resulting
distribution is highly peaked at probabilities 0 and 1, such that $prob >$ 0.8
results in a highly pure sample of galaxies. We then searched the resultant
catalog of galaxies for systems that are likely to be clusters of galaxies
using the \maxbcg\ algorithm described briefly below and in detail in
\citet{Koester07a}.

For characterizing the galaxy population in clusters, we restrict the galaxy
sample to be volume and magnitude limited for $z \le 0.3$ by including only
those galaxies brighter than an appropriate threshold.  Expressed as a
luminosity K-corrected to redshift $z=0.25$ (see \S\ \ref{sec:kcorr}), this lower
$i$-band luminosity limit is $10^{9.5} h^{-2} L_{\odot}$. At $z=0.3$, this
value corresponds to an apparent magnitude limit of $i < 21.3$, and color
limits of \gmr\ $< 2$ and $^{0.25}(r-i) < 1$. All magnitudes are SDSS model
magnitudes. Uncertainties on the measurements are sufficiently small so that
the typical $r-i$ signal-to-noise is greater than 30 until beyond redshift 0.3. Galaxy
colors are therefore robustly measured \citep{Smail95b, Bernstein02, Benitez04} over the
full redshift range considered here.

\begin{figure*}  
  \epsscale{0.95} 
  \plotone{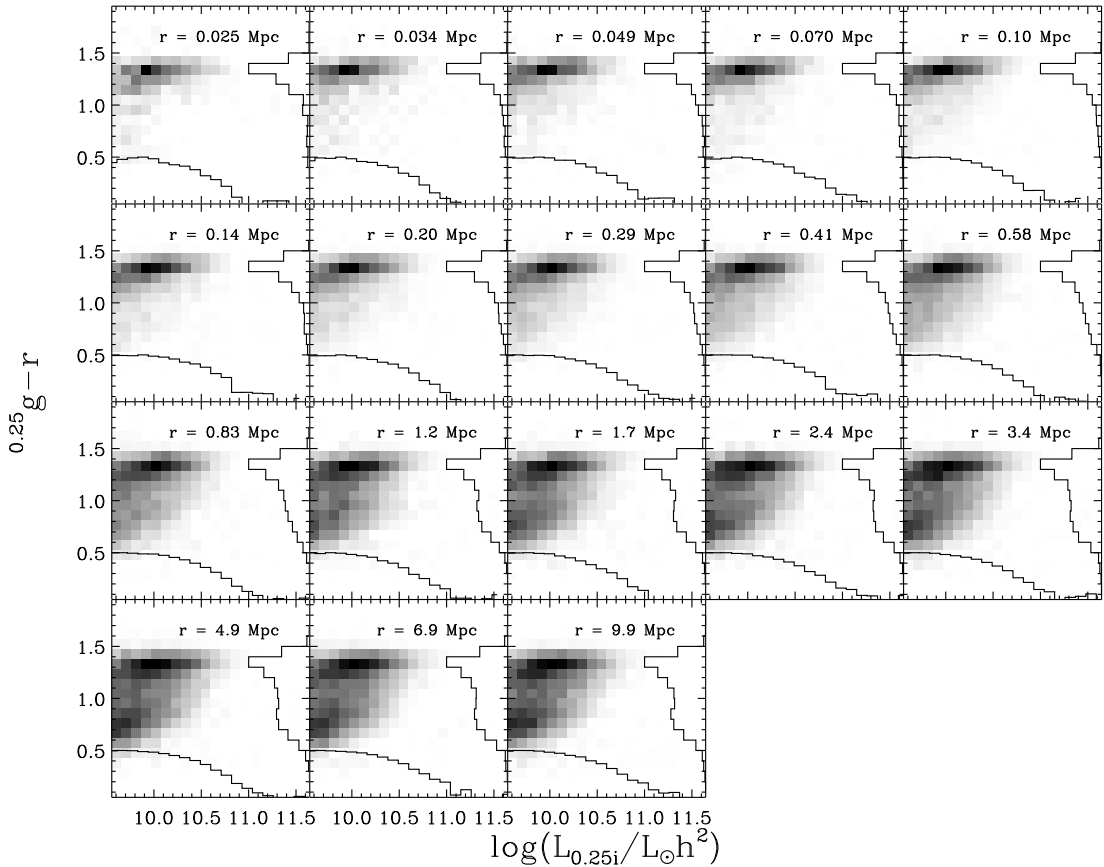} \figcaption[example_col_lum.eps]{Joint
    distribution of \gmr\ and $^{0.25}i$-band luminosity density as a
    function of projected separation from cluster centers for an
    example richness bin, 12 $\le$ \Ntwo\ $\le$ 17.  Each panel
    corresponds to a different radial bin as indicated.  The
    one-dimensional distributions for luminosity and color are shown
    along the $x$- and $y$-axes respectively. The luminosity
    distribution is expressed as log of the number density as a
    function of log luminosity; the color distribution is given as the
    linear density as a function of color.  At small separations, red
    galaxies dominate, while on large scales there is a bivariate
    color distribution similar to the global distribution.  A
    smaller fraction of galaxies are highly luminous near cluster
    centers.
\label{fig:lumcolor_vs_rad}
}
\end{figure*}

\subsection{Cluster Sample} \label{sec:clustersamp} 

Clusters were identified using the \maxbcg\ cluster finder of
\citet{Koester07a}.  This algorithm identifies clusters by the presence of a
BCG and a red sequence, and provides an excellent photometric redshift  for
each system.  The scatter in redshift is $\Delta z \lesssim 0.02$ for our full
richness and redshift range and $\Delta z \sim 0.015$ for \Ntwo\ $\ge 10$.  The
cluster center is defined to be the position of the BCG. The cluster
richness, \Ntwo, is defined as the number of red galaxies with rest frame
$i$-band luminosity $L > 0.4$\Lstar\ in an aperture that scales with \rtwo, and
is an excellent mass proxy, as discussed below. The catalog of systems with
\Ntwo\ $\ge$ 10 was presented in \citet{Koester07b}; in this work we use systems
with \Ntwo\ $\ge$ 3.

The \maxbcg\ cluster finder was trained using known BCGs and luminous red galaxies, and as
we will consider whether the resulting BCG population is being constrained by
these priors, we review the procedure here.  Specifically, when constructing the likelihood of a galaxy to be
a BCG, the algorithm places priors on observed $g-r$ and $r-i$
colors and $i$-band magnitude.  The model BCG color--redshift and
magnitude--redshift relationships are taken from the distribution of bright LRG
galaxies, then refined based on examination of 100 visually identified BCGs in
rich clusters. The BCG likelihood, $\mathcal{L}_{BCG}$, is specified to be
 \begin{equation}
 \mathcal{L}_{BCG}(z)=G_{g-r}^{BCG}(z)\, G_{r-i}^{BCG}(z)\, \mathrm{e}^{-((m-m_i)/\sigma_c)^2}
 \end{equation}
\citep[][Equation 11]{Koester07a}, where the width $\sigma_c$ of the
prior on BCG $i$-band magnitude is 0.3mag, taken from \citet{Loh06}.
In addition, BCGs are required to be brighter than 0.4\Lstar\ ($6.9
\times 10^9 h^{-2} L_\odot$ at $z = 0.25$). The distributions
$G_{g-r}^{BCG}$ and $G_{r-i}^{BCG}$ are narrow Gaussians in color; the
width of each includes both the intrinsic width of the E/S0 ridgeline
(0.05 for $g-r$ and 0.06 for $r-i$) and the error in the measured
color for the galaxy in question. In practice, the narrow color priors
provide a much stronger constraint in identifying likely BCGs than
does the magnitude prior, and for the majority of cases, the BCG is
clearly defined by the color criteria.  The BCG magnitude likelihood
has an appreciable effect only for the cases where the choice of BCG
is ambiguous.

The selection function of the \maxbcg\ algorithm was investigated by
\citet{Koester07a} and \citet{Rozo07a}, who showed that this sample of
clusters is over 90\% complete and pure for \Ntwo\ $\ge 10$. For
\Ntwo\ $\ge 20$ sytems, the completeness and purity rates are 97\% or
better.
The completeness and purity were quantified by
comparing clusters identified in realistic mock galaxy catalogs
\citep{Wechsler} to their host halos.  We did not wish to use unresolved halos
in our comparisons, so for lower richness values (\Ntwo\ $< 10$), the completeness and purity
are less well quantified.  It is likely that the lowest-richness systems are
less pure and complete, not only because of statistical considerations but also
because the lower richness systems, with just a few red galaxies in close
proximity, are not necessarily representative of all systems in the equivalent
mass range.  Higher resolution simulations will be necessary to quantify this
selection in detail; however, we can learn about the statistical properties of
these systems even if they are a biased population compared to dark matter
halos.  We thus include them in this study, but for the purpose of fitting
relationships of galaxy properties as a function of cluster richness or mass,
we restrict the sample to \Ntwo\ $\ge 10$, where the catalog is known to be both
highly complete and pure.

We use clusters in the redshift range 0.1 $\le z \le$ 0.3; there are
\nclust\ objects identified with \Ntwo\ $\ge$ 3, of which 13,823 have
\Ntwo\ $\ge 10$.

\subsection{Average Galaxy Properties at $z = 0.25$}

Below, we compare our derived cluster galaxy population statistics to similar
statistics for the mean of galaxies in all environments, which we will refer to as the
cosmological, or universal, average. To calculate these average galaxy
statistics, we use the New York University Value-Added Galaxy Catalog
\citep[NYU-VAGC][]{Blanton05a} of the SDSS spectroscopic sample, whose median
redshift is $z \sim 0.1$, evolved to $z = 0.25$. The details of constructing
this sample are discussed in \citet{Sheldon07b}. By applying the same
luminosity and color cuts to this sample as we do to any given set of cluster
galaxies (\eg, to determine the density of red, bright galaxies), we
estimate the universal average values of a comparable ``field'' sample. We do
not undertake a detailed study of this population here, but just present the
average values for general comparison with the recovered cluster values. For
reference, from fitting a Schechter function to this sample, we find the global
value of $L_*$ evolved to $z=0.25$ to be $1.7 \times 10^{10}h^{-2}L_\odot$.

\section{Method} \label{sec:methods} 

In this section, we describe how we characterize the radial, luminosity, and
color distribution of galaxies that are associated with clusters. In \S\
\ref{sec:estimator} we review the estimator used to determine the distribution
of cluster galaxies; \S\S\ \ref{sec:randoms} through \ref{sec:mass-obs} discuss
the technical details of characterizing the survey geometry and redshift
distributions, performing K-corrections, and determining \Mtwo\ and \rtwo. We
describe a check to the background correction with a mock galaxy catalog in \S\
\ref{sec:bkgtest}.

We can only reliably determine cluster membership for individual
clusters in this purely photometric sample for red, relatively high
luminosity galaxies.  We nonetheless can accurately characterize the
full cluster galaxy population by statistically correcting for
galaxies aligned by chance along the line of sight.  We perform this
correction not on individual clusters, but for ensembles binned
together based on their observable properties, such as richness or
redshift.  We do not recover the properties of individual clusters,
but rather properties of the average cluster.  In this sense we
measure the cross-correlation between clusters and the galaxy
population.

If the average stacked cluster is spherically symmetric, we can invert the
projected densities to recover the 3D profile.  This assumption is true
for a homogeneous and isotropoic universe, as long as the cluster finder
does not introduce anisotropies: for example, choosing preferentially
systems elongated along the line of sight.

To recover the K-corrected luminosities of the average galaxy population, we
K-correct galaxies along both the cluster line-of-sight and the random
lines-of-sight to the redshift of the cluster, regardless of their true
redshifts.  After background subtraction, galaxies at redshifts different
from the cluster are removed from the counts, and so this inaccuracy does
not persist in our final measurements.  

A correlation function-based approach is straightforward to implement and
interpret, and takes advantage of the full dataset for background correction.
This method is in contrast to using a ``local'' background correction estimate
from, for example, counting galaxies in a large annulus around each cluster, as
has been done in studies of more limited area. As there is mass and light
correlated with clusters out to many times the virial radius, a local
background estimate is likely to result in a biased estimate of the true galaxy
population in clusters. \citet{Masjedi06a} implemented the use of a correlation
function-based estimator to investigate galaxies correlated with the SDSS LRG
sample, and here we extend this technique to examine galaxies associated
\maxbcg\ clusters.  This estimator is a powerful tool, but requires a
significant amount of computation for the large number (here \altnumpairsTenMpc)
of pairs that must be considered.

\subsection{Cluster-galaxy Correlations} \label{sec:estimator} 

We follow the method of \citet{Masjedi06a} to estimate the mean number density
and luminosity density of galaxies associated with clusters.  This method
essentially calculates a correlation function with units of density; it
includes corrections for random pairs along the line of sight as well as pairs
missed due to edges and holes. In the description of the estimator below, we
discuss the necessary terms for calculating the luminosity density. The number
density measurement is simply a special case where we weight each galaxy by
unity rather than by its luminosity. 

We define two samples: the primary sample, denoted $p$, and the secondary
sample, denoted $s$, that are either in the real data $D_x$ or random locations
$R_x$.  For example, the counts of real data secondaries around real data
primaries is denoted $D_p D_s$, while the counts of real data secondaries
around random primaries is $R_p D_s$.  For this study, the real data primaries
are galaxy clusters with redshift estimates and the secondaries are the imaging
sample of galaxies with no redshift information. The generation of random data is discussed in \S\ \ref{sec:randoms}.

The mean excess luminosity density of secondaries associated with primaries is 
\begin{equation} \label{eq:estimator}
\bar{\ell}~ w = \frac{D_p D_s}{D_p R_s} - \frac{R_p D_s}{R_p R_s},
\end{equation}
where $\bar{\ell}$ is the mean luminosity density of the secondary sample,
averaged over the redshift distribution of the primaries, and $w$ is the
projected correlation function. This quantity is the estimator from \citet{Masjedi06a}
where the weight of each primary-secondary pair is the luminosity of the
secondary.  We have written the measurement as $\bar{\ell} w$ to illustrate that
the estimator gives the mean density of the secondaries times the projected
correlation function $w$. Only the excess density with respect to the mean can
be measured with this technique.  Using a weight of unity for each galaxy,
rather than luminoisty, gives the mean number density.

The first term in equation \ref{eq:estimator} estimates the total luminosity
density around clusters, including everything from the secondary imaging sample
that is projected along the line of sight. The second term estimates the
contribution from random objects along the line of sight.  Note, the same
secondary may be counted around multiple primaries (or random primaries).

The numerator of the first term, $D_p D_s$, is calculated as
\begin{equation}
D_p D_s = \frac{\sum_{p,s}{L_s}}{N_p} 
  = \langle L_{pair} \rangle + \langle L^R_{pair} \rangle 
  = f A~\bar{\ell}~(w + 1),
\end{equation}
where the sum is over all pairs of primaries and secondaries, weighted by the
luminosity of the secondary.  The secondary luminosity is calculated by
K-correcting each secondary galaxy's flux assuming it is at the same redshift
as the primary (see \S\ \ref{sec:kcorr} for details). The total luminosity is
the sum over correlated pairs ($L_{pair}$) as well as random pairs along the
line of sight ($L^R_{pair}$). By the definition of $w$, this number is the
total luminosity per primary times $w+1$. This term can be rewritten in terms
of the luminosity density of the secondaries $\bar{\ell}$ times the area probed
$A$.  Some fraction of the area searched around the primaries is empty of
secondary galaxies due to survey edges and holes. The geometry factor $f$
represents the mean fraction of area around each primary actually covered by
the secondary catalog. The geometry factor is a function of pair separation,
with a mean value close to unity on small scales but then dropping rapidly at
large scales. Measurement of the survey geometry is discussed in \S\
\ref{sec:geom}.

The denominator of the first term calculates the factor $f A$, the actual area
probed around the primaries.  This term in the denominator corrects for the
effects of edges and holes. Also, because the denominator has units of area, we
recover a volume density rather than just the correlation function. This term is
calculated as:
\begin{equation}
D_p R_s = \frac{ N^{DR}_{pair} }{ \sum_{p} \left( \frac{d\Omega}{dA} \right)_p \frac{dN}{d\Omega}  }
  = f A.
\end{equation}
The numerator is the pair counts between primaries and random secondaries, and
the denominator is the expected density of pairs averaged over the redshift
distribution of the primaries, times the number of primaries. The ratio is the
actual mean area used around each primary $f A$.

The second term in equation \ref{eq:estimator} accounts for the random pairs
along the line of sight.  The numerator and denominator of this term are
calculated in the same way as the first term in equation \ref{eq:estimator},
but with randomly chosen locations as primaries distributed over the survey
geometry.  The redshifts are chosen such that the distribution of redshifts
smoothed in bins of $\Delta z = 0.01$ match that of the clusters.  The 
numerator and denominator are
\begin{eqnarray}
R_p D_s = \frac{\sum_{pr,s}{L_s}}{N_p} 
  = \langle L^R_{pair} \rangle 
  = f^R A^R~\bar{\ell}
\end{eqnarray}
and
\begin{eqnarray}
R_p R_s = \frac{ N^{RR}_{pair} }{ \sum_{pr} \left( \frac{d\Omega}{dA} \right)_{pr} \frac{dN}{d\Omega}  }
  = f^R A^R,
\end{eqnarray}
respectively. The ratio of these two terms, $R_p D_s / R_p R_s$, calculates the
mean density of the secondaries around random primaries after correcting for
the survey geometry.

\begin{figure}
  \epsscale{1.15} 
  \plotone{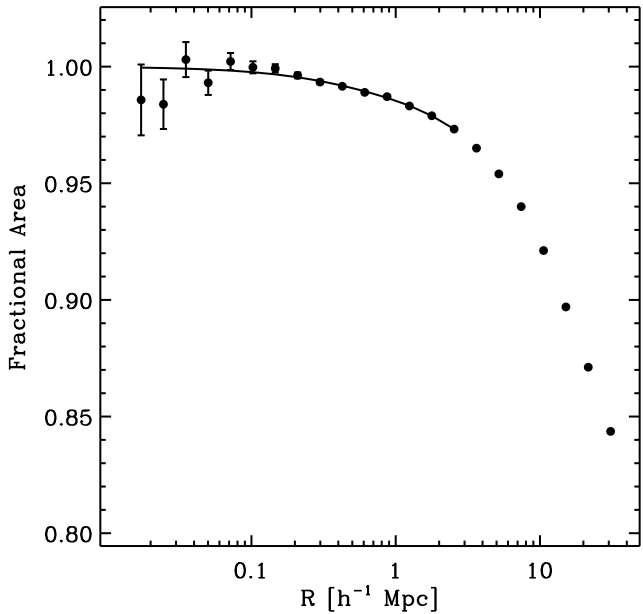}
  \figcaption[example_usedarea.eps]{Mean fractional area searched
    relative to the area in the radial bin for one bin in cluster
    richness.  Edges and holes dilute the true galaxy counts, biasing
    the measured density.  This effect is negligible on small scales,
    but becomes important for large distances from clusters, where a
    higher fraction of clusters are closer to the edge than the search
    radius.  Due to the small area enclosed at small separations, the
    value is not well determined, but must approach unity smoothly.
    We model this behavior with a polynomial constrained to unity on small
    scales.  For larger scales no model is
    needed. \label{fig:usedarea}}  \vspace{1em} 
\end{figure}

The density measured with this technique may be tabulated in various ways,
typically as a function of projected radial separation $R$.  We tabulate in a
cube that represents bins of separation $R$, luminosity $^{0.25}L_i$, and
color \gmr.  This choice facilitates the study of the radial dependence of the
luminosity function and the color--density relation. We make measurements for
objects in the ranges $0.02 < R < 11.5 h^{-1}$ Mpc (in 18 bins), $0 <$ \gmr\ $<
2$ (in 20 bins) and $9.5 < $ log$_{10}(^{0.25}L_i/L_{\odot}) < 11.7$ (in 20
bins), where $g-r$ and $L_i$ were K-corrected to the median redshift of the
cluster sample, $z = 0.25$. Note that the smallest separation that we use is
20$h^{-1}$ kpc, interior to which the light is dominated by the BCG. We retain
information about the BCG of each cluster separately. An example of the
resulting distributions is shown in Figure \ref{fig:lumcolor_vs_rad}: each
panel shows the joint color--luminosity distribution in the richness bin 12
$\le$ \Ntwo\ $\le$ 17 for a different radial range, the mean of which is
indicated in the legend of each panel. We save the cubes of luminosity density
and number density data for every cluster, then bin the clusters into samples
based on cluster observables. We evaluate the uncertainties for our results
using jackknife estimation from dividing the survey area into 12 roughly
contiguous, roughly equal area sections.

For each bin in richness, the projected radial profiles as a function of
luminosity and color are inverted using a standard Abel type integral to
recover the three-dimensional profiles. The inversion relates the derivative of
$N (R)$, the projected number density profile, to $n (r)$, the 3D density
profile, as
\begin{equation}
\Delta n(r) = n(r) - \bar n = \frac{1}{\pi} \int_{r}^{\infty}dR\frac{-N^\prime (R)}{\sqrt{R^2 - r^2}} 
\end{equation}
where $\Delta n (r)$ is the excess density over the mean random background
$\bar n$. Inherent in this inversion is the assumption that $N (R)$ is the
line-of-sight projection of a spherically symmetric $n (r)$. In an isotropic
universe this assumption is valid as long as cluster selection does not
preferentially find systems oriented in a particular direction with respect
to the observer.

We apply the nonparametric method of \citet{Johnston07a} to deproject the
number density profiles.  Because the maximum separation we use is 10\Mpc, not
infinity, we cannot constrain the 3D density at the outermost point, and the
second to last point, at $\sim 8$\Mpc, must be corrected. With a power law
extrapolation of the profile, the correction is $\sim$ 5\% upward for that
point. Deprojection is necessary to accurately recover the the red fraction and
luminosity functions. Failure to properly deproject will result in artificially
high faint-end LF slopes and artificially low red fractions
\citep{Valotto01,Barkhouse07}. After deprojecting, we translate the bins of physical radial
separation into bins of \rad\ using the measured \rtwo-\Ntwo\ relationship (see
\S\ \ref{sec:mass-obs}). To make measurements within the same bins of \rad\ for
clusters of different richness, we use power-law interpolation between the
fixed $R$ values tabulated in the data cube.

We construct the LF of clusters in a given richness range in the following
manner.  For each richness value, for a given bin in luminosity, we calculate
the excess number density of galaxies in the desired color and \rad\ range.
For the LF of a broader richness bin, we average the LFs of all clusters in
that richness range. To measure the red fraction for clusters in a given
richness range, we follow an analogous procedure: we find the number of red and
all galaxies in the desired luminosity and \rad\ range for each cluster
richness, then take the average over all clusters within the broader richness
bin. The number of red galaxies divided by the total number of galaxies (within
the luminosity and \rad\ limits) is the red fraction, \fred.  To measure the
total luminosity of galaxies in a given \rad, color, luminosity and richness
range, we again follow the same procedure, but using the tabulated values of
luminosity density, rather than number density.

These measurements allow investigation of the distribution of
excess-over-random galaxies around the BCGs used as \maxbcg\ centers. If a
significant fraction of BCGs are miscentered relative to the density peak of
the dark matter halos (whether because the true BCG does not lie at the deepest
part of the potential well or because the wrong galaxy was identified as the
BCG during cluster detection), the resulting weak lensing profiles can be
affected \citep{Johnston07a}, as should the light profiles. In this work we
make no miscentering correction, but remind the reader that we measure the
distribution of galaxies around BCGs, not halo mass peaks.

\subsection{Random Catalogs} \label{sec:randoms} 

To correct for the contribution of galaxies along the lines of sight to
clusters, we generate a set of 15 million random points. These points are used
in the $R_pD_s$, $D_pR_s$ and $R_pR_s$ terms from the estimator described in
equation \ref{eq:estimator}. The random positions are distributed uniformly
over the survey area using the window function described in \S\ \ref{sec:geom}.
The redshifts used for random points must statistically match that of each
cluster sample in order for the background subtraction to be accurate. To
achieve the proper distribution, the random primaries were generated with
constant comoving density.  For a given subsample of clusters we re-weight the
random primaries so that the weighted redshift distribution matches that of
each cluster subsample when binned with $\Delta z$ = 0.01.

\subsection{Survey Geometry} \label{sec:geom}

We characterize the survey geometry using the SDSSPix
code\footnote{\tt http://lahmu.phyast.pitt.edu/$\sim$scranton/SDSSPix/}. This code
represents the survey using nearly equal area pixels, including edges and holes
from missing fields and ``bad'' areas near bright stars.  We remove areas with
extinction greater than 0.2 magnitudes in the dust maps of \citet{Schlegel98}.
This window function was used in the cluster finding and in defining the galaxy
catalog for the cross-correlations.  By including objects only from within the
window, and generating random catalogs in the same regions, we control and
correct for edges and holes in the observed counts as described in \S\
\ref{sec:estimator}. The resulting area examined is 7398.23 deg$^2$.

The denominator terms $D_pR_s$ and $R_pR_s$ in equation
\ref{eq:estimator} correct for the survey edges and holes by measuring
the actual area searched.  An example $D_pR_s$ is shown in Figure
\ref{fig:usedarea}, generated for one richness bin. This quantity is
expressed as the mean fractional area searched relative to the area in
the bin. For small separations, edges and holes make little
difference, so the fractional area searched is close to unity, but on
larger scales edges become important.

On very small scales the area probed in each bin is relatively small
and the correction factor is not well constrained.  However, we know
that the area correction factor must be close to unity, a fact that is
clear from visual inspection.  To smooth the result, we fit a fifth
order polynomial, constrained to be unity on small scales, to the
fractional area as a function of the logarithmic separation.  Due to
the weighting, this approach results in a curve that approaches unity
smoothly on small scales, yet matches intermediate separation points
exactly.  Points on larger scales are well-constrained and do not need
smoothing.

\subsection{K-corrections} \label{sec:kcorr}

We calculated K-corrections using the template code \texttt{kcorrect}
from \citet{BlantonKcorr03}.  This code is accurate but too slow to
calculate the K-corrections for the over $10^9$ pairs found in
the cross-correlations. To make the computation tractable, we computed
 K-corrections on a grid of colors in advance using galaxies from
the SDSS Main sample as representative of all galaxy types. We
computed the K-corrections for these galaxies on a grid of redshifts
between 0 and 0.3, the largest redshift considered for clusters in
this study.  The mean K-correction on a 21x21x80 grid of observed
$g-r, r-i$, and $z$ was saved.  We interpolated in this cube when
calculating the K-correction for a given galaxy. This interpolation
makes the calculation computationally feasible for this study, but is
still the bottleneck.

To minimize uncertainties in the K-correction, we K-correct to the
median redshift of the cluster sample, $z = 0.25$. All reported
magnitudes are adjusted to this redshift, and are noted as \eg,
$^{0.25}M_i$. \citet{Blanton03} give a complete discussion of this
bandpass shifting procedure.

\subsection{r$_{200}$ and M$_{200}$} \label{sec:mass-obs}

\begin{figure}
  \epsscale{1.15} \plotone{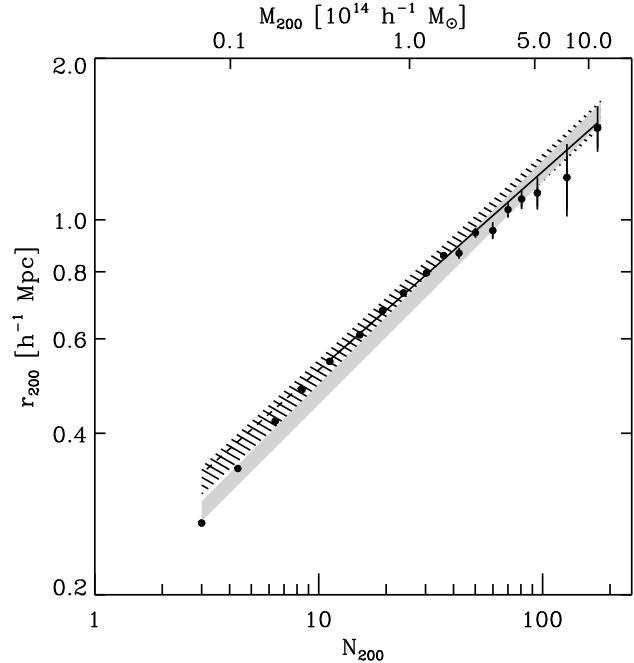}
  \figcaption[r200_alsobecker.eps]{Cluster radius \rtwo\ as a function
    of cluster richness, measured with three different methods.  The
    result from this paper, based on the distribution of galaxies, is
    shown as the points with error bars; the solid line is the best
    fit power law to systems with \Ntwo\ $\ge$ 10 and the dashed line extends this relationship to lower richness.  These data are compared
    with results from velocity dispersion profiles (hashed region;
    \citealt{Becker07}) and from weak lensing (shaded region;
    \citealt{Johnston07b}). The error bars on the galaxy data
    points are determined from jackknife resampling of the data. The
    width of velocity dispersion-based estimate corresponds to an 18\%
    uncertainty in mass; the width of the lensing region results from
    the reported 13\% uncertainty in mass.
    \label{fig:r200ngals}}
  \vspace{1em}
\end{figure}

There has been much recent work to quantify the mass--observable
relation for stacked samples of the maxBCG clusters, using mass
estimates from cluster abundance \citep{Rozo07b}, velocity dispersion
\citep{Becker07}, weak lensing \citep{Sheldon07a, Johnston07b}, and
X-ray measurements \citep{Rykoff07}. These various methods all result
in a consistent mass--richness scaling; a detailed comparison of the
different mass estimators is in progress (Rozo et al. in
preparation). These measurements also yield estimates of cluster size.

\begin{figure}[t]
  \epsscale{1.2} \plotone{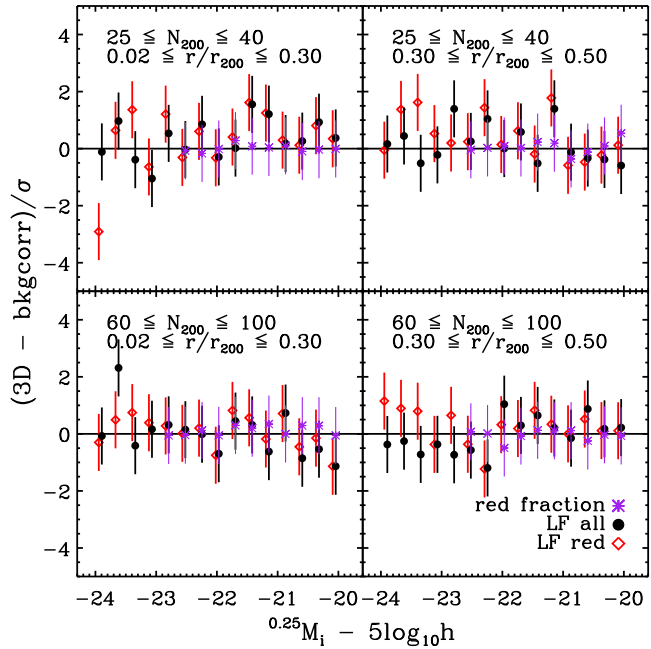}
    \figcaption[bkgtest_new.eps]{A test of the background-correction
    algorithm with the ADDGALS mock catalog. Each panel shows the
    difference between the underlying 3D distribution of luminosity
    and the background-correction algorithm's recovered distribution,
    in units of the uncertainty $\sigma$. The {\bf left} column shows
    galaxies within $0.3\times$\rtwo\ of cluster centers, while the {\bf right}
    column shows galaxies in the radial range 0.3 $\le$ \rad\ $\le$
    0.5. The {\bf top} panels use galaxies around
    clusters in a low richness bin (25 $\le$ \Ntwo\ $\le$ 40), and the
    {\bf bottom} panels use clusters in a high richness bin (60 $\le$
    \Ntwo\ $\le$ 100). In each panel the difference, in units of
    $\sigma$, is shown for the {\it red fraction} (purple asterisks),
    the density of {\it all} galaxies (black, filled circles) and of
    {\it red} galaxies (red, open diamonds). For magnitudes brighter
    than \Mi\ $\sim$ -23, no galaxies are found, and so the red
    fraction is not calculated. In each case the background-correction
    algorithm accurately recovers the underlying distribution.
  \label{fig:bkgtest}} 
  \vspace{1em}
\end{figure}

To compare equivalent regions of clusters of various masses and thus
various sizes, we scale all physical radii to \rtwo, where \rtwo\ is
the threshold radius interior to which the mean mass density of a
cluster is 200 times the critical mass density of the Universe. The
weak lensing-measured \rtwo\ of \citet{Johnston07b} yields a size-richness relationship of
\begin{equation}
r_{200} = 0.182 h^{-1} {\rm Mpc}\, N_{200}^{0.42},
\end{equation}
which we use here. As we have measured 3D profiles around all clusters
out to 8\Mpc, we can construct 3D radial profiles out to 5$\times$\rtwo\ for
even the most massive systems in our catalog.

\begin{figure}[t]
  \epsscale{1.2} \plotone{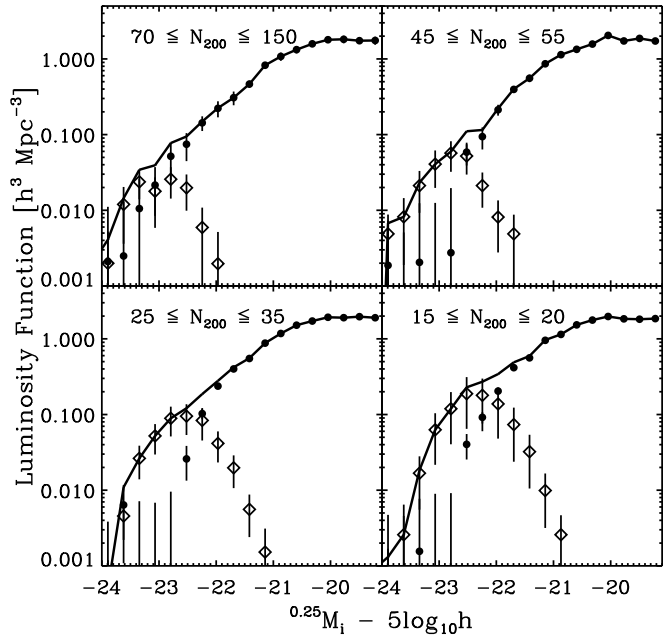}
  \figcaption[example_lfs.eps]{Luminosity functions of galaxies within
    \rtwo\ for four example bins in cluster richness, showing the
    contribution from BCGs (open diamonds) and satellite galaxies
    (filled circles) to the total LF (solid line).  The figure
    illustrates one way in which the BCG population is different from
    the satellite distribution. Error bars are from jackknife
    resampling, and are omitted on the total distribution
    (BCG+satellites) for clarity. \label{fig:LFwithBCG}}
  \vspace{1em}
\end{figure}

For comparison, we also derive \rtwo\ using the approach presented in
\citet{Hansen05} for estimating \rtwo\ as a function of \Ntwo\ using
the space density of galaxies in clusters. In that work, the radial
space density profile of cluster-correlated galaxies brighter than a
luminosity threshold was compared with the global mean space density
as determined from the global SDSS luminosity function
\citep{Blanton03}, evolved to $z = 0.25$, integrated to the same luminosity threshold. The
radius interior to which the mean space density of cluster galaxies
reached 200/$\Omega_M$ times the global mean density of galaxies was
taken as \rtwo. If galaxies were completely unbiased with respect to
dark matter on all scales, \rtwo\ measured with galaxies should
exactly match \rtwo\ measured from the true mass distribution. The
characteristic cluster-associated galaxy used has a luminosity of
slightly sub-\Lstar. On average in the Universe, a galaxy with this
luminosity has a bias of close to unity. As our sample is specifically
chosen to be galaxies associated with clusters, however, and clusters
do not present the same environment as the cosmological average, we
cannot necessarily expect that the typical cluster galaxy thus also
has bias close to one. Nonetheless, we show below that using these
galaxies as a tracer of the total matter distribution does result in a
reasonable approximation of \rtwo, a finding also noted by
\citet{Gonzalez07}.

\begin{figure*}
  \plottwo{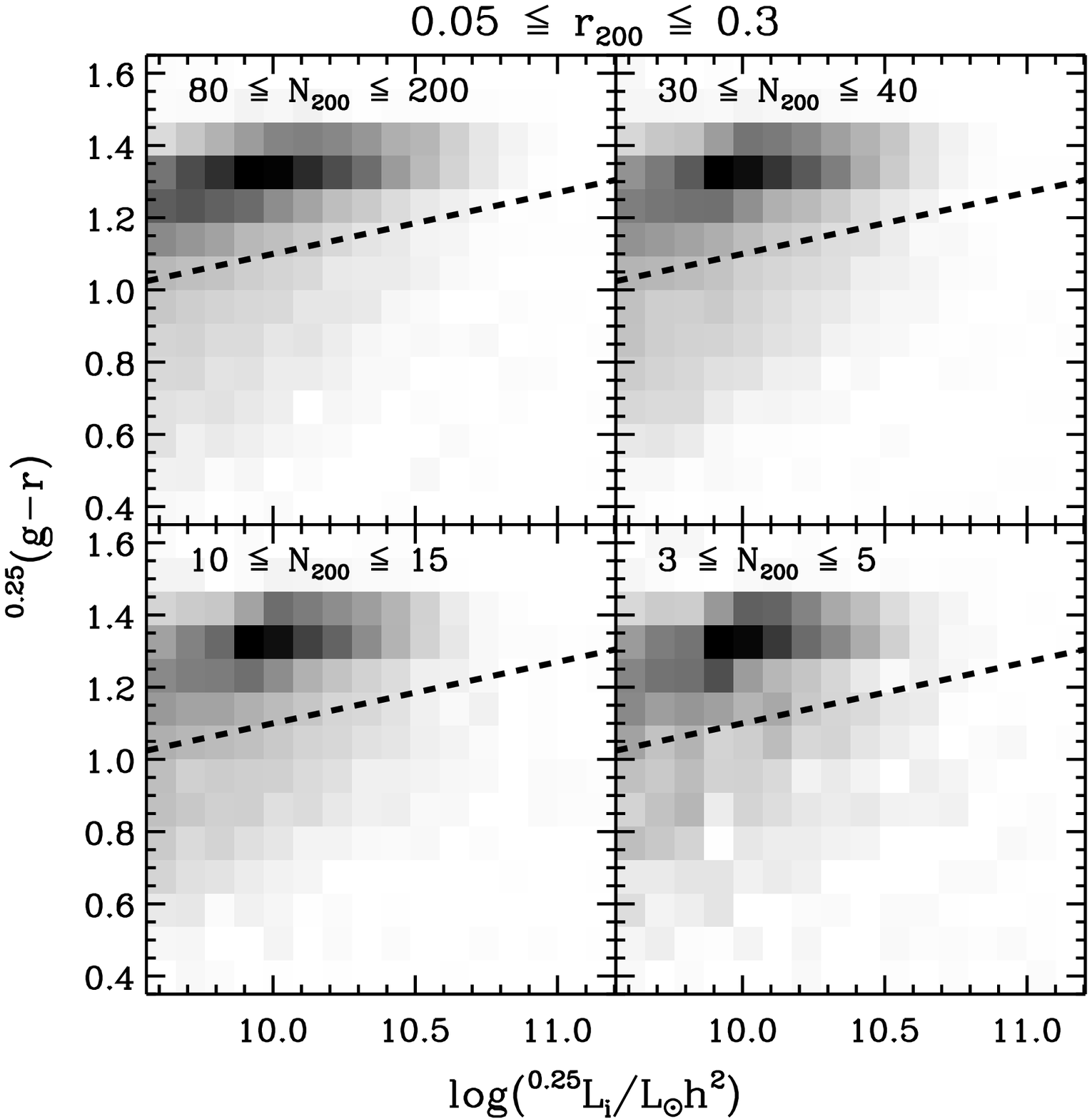}{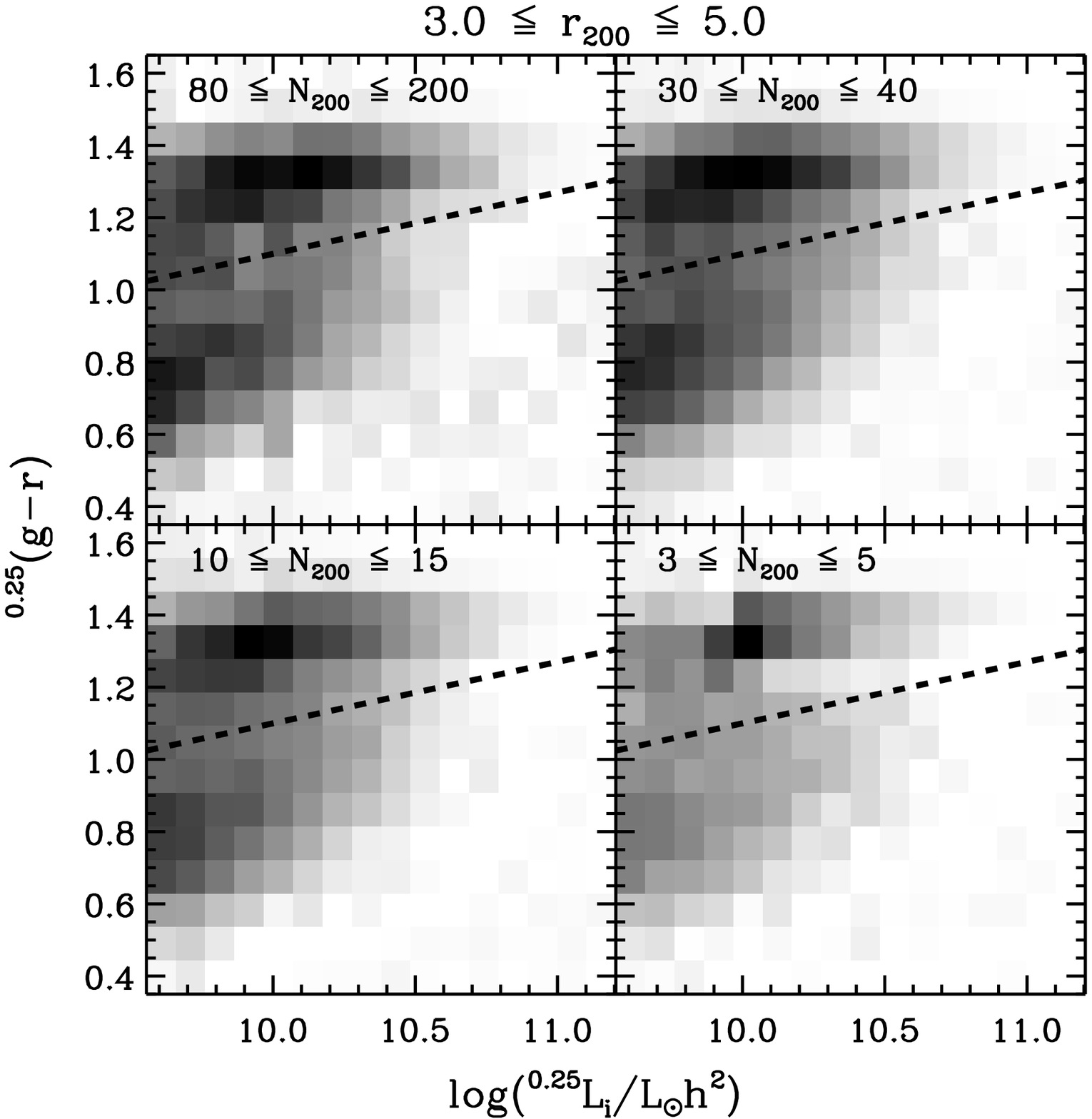}
  \figcaption[cm_inner.eps]{Color--luminosity distributions for
    satellite galaxies in clusters, in several example richness and
    radial ranges.  Shading shows the relative density in
    color--luminosity space for each bin.  {\bf Left:} galaxies with
    $0.05 \le r/$\rtwo\ $ \le 0.3$; {\bf right:} galaxies in the range
    $3 \le r/$\rtwo\ $ \le 5$. The dashed line indicates the (fixed)
    split used to separate galaxies into red and blue samples, given
    by equation \ref{eq:colsplit}.\label{fig:cmsplit}}
\end{figure*}

Using the galaxy distribution-based method with the current data set,
and comparing to the $z = 0.25$ field sample, we find for \Ntwo\ $\ge
10$ that \rtwo\ $= (0.224 \pm 0.004) h^{-1}$ Mpc $N_{200}^{0.37 \pm
  0.01}$. Figure \ref{fig:r200ngals} shows \rtwo\ as estimated from
the galaxy distribution (points with error bars), the relationship
from the velocity dispersion-based mass estimate (hashed region), and
the lensing estimate (shaded region). The width of the hashed region
corresponds to a 18\% uncertainty in the mean mass--observable
relationship from \citet{Becker07}; the width of the shaded region
corresponds to a 13\% uncertainty in mass as reported for the lensing
analysis in \citet{Johnston07b}. The best fitting power law for \Ntwo\
\gae\ 10 for the galaxy-based measurement is plotted as the solid line,
with a dashed line extending the trend to lower richness. The \rtwo\
derived from the galaxy distribution is remarkably consistent with
these other observations. The slight discrepancy in scaling gives
hints about the way that galaxies populate halos of dark matter, but
further investigation of this interesting problem is beyond the
intended scope of this work.

The measurement of \rtwo\ also enables an estimate of cluster mass,
\Mtwo. For the present work, we present our results as a function of
the direct observable, \Ntwo, but use the \rtwo--\Ntwo\ relationship
from the weak lensing analysis to translate this observable into a
mass estimate. The mass is calculated from \rtwo\ via
\begin{equation}
M_{200} = 200 \rho_c(z)\frac{4}{3} \pi r^3_{200} 
\end{equation}
where $\rho_c(z) = 3H^2(z)/(8\pi G)$ is the critical density at epoch
$z$, and the Hubble parameter is given by $H^2(z) =
H^2_0[\Omega_m(1+z)^3 + (1-\Omega_m)]$ for a flat LCDM Universe. For
the choices of $z=0.25$ and $\Omega_m=0.27$ used in
\citet{Johnston07b}, the $M_{200}$--\rtwo\ conversion is $M_{200} =
2.923 \times 10^{14} h^{-1}M_{\odot} (r_{200}/h^{-1}{\rm Mpc})^3$.
We therefore adopt
\begin{equation}
M_{200} = 1.75 \times 10^{12} h^{-1} M_{\odot} \, N_{200}^{1.25}, 
\end{equation}
the best-fit value from \citet{Johnston07b}. We note that these values
are measured in the lensing data including the effects of
miscentering, and so should reflect the underlying halo mass, rather
than the mass correlated with cluster centers.

\subsection{Testing with Simulation Data} \label{sec:bkgtest}

To check that our methods of background correction and deprojection
are reliable, we compare the SDSS data to results from a mock catalog
from an N-body simulation populated with galaxies using the ADDGALS
method of of \citet{Wechsler}.  This catalog is derived from the
N-body Hubble Volume light-cone simulation \citep{Evrard02}. Particles
in the simulation are assigned luminosities to match the
luminosity-dependent two-point correlation function measured in the
SDSS by \citet{Zehavi04} and the global SDSS luminosity function
\citep{Blanton03}. Colors are assigned depending on local environment
as defined by the distance to the 5th nearest neighbor, matching the
photometric properties of SDSS galaxies with similar luminosities and
local densities. The luminosity limit of the mock catalog is L $>$
0.4\Lstar, matched to the luminosity threshold used for identifying
\maxbcg\ clusters.  This mock catalog successfully reproduces the
width, location and evolution of the red ridgeline, making it ideal
for testing and understanding the selection function of the \maxbcg\
algorithm.  A detailed characterization of the \maxbcg\ selection
function based on this catalog has been undertaken by
\citet{Koester07a} and \citet{Rozo07a}.

We use this simulation to test the algorithms presented for
quantifying the galaxy population.  We run the \maxbcg\ cluster finder
on the mock catalog, then implement the same background-correction
algorithm used on the SDSS data to characterize the galaxy population
statistically associated the \maxbcg-identified cluster centers.  In
order to test the algorithms, we compare the properties of galaxies
located in three dimensions around \maxbcg-identified cluster centers
with the statistical distribution of galaxies determined by the
background-correction algorithm run on the mock.  The mass limit for
resolved halos in the simulation, $M_{200} = 5 \times 10^{13}$ \Msun,
allows a close to complete sample for a richness of \Ntwo\ $\ge 10$,
so we focus on results above that range.

We find that the background-correction algorithm does well at
reproducing the underlying distribution. For example, Figure
\ref{fig:bkgtest} demonstrates that the difference between the
underlying 3D values and the background-corrected values for several
statistics are typically within $\sim 1 \sigma$.  Shown is the
comparison between the luminosity functions of all galaxies (filled
circles) and red galaxies (diamonds). Also shown is the comparison of
red fraction values (asterisks). In all cases the difference is
expressed in units of the uncertainty $\sigma$. The left column is for
galaxies within $0.3\times$\rtwo, while the right column is for galaxies in
the range $0.3 \le$ \rad\ $\le 0.5$; the top panels are a low richness
bin (25 $\le$ \Ntwo\ $\le$ 40) and the bottom panels are a high
richness bin (60 $\le$ \Ntwo\ $\le$ 100).  For magnitudes brighter
than \Mi\ $\sim$ -23, there are no galaxies found, so the red fraction
is not calculated.

\begin{figure*}
   \epsscale{1.2}
   \plotone{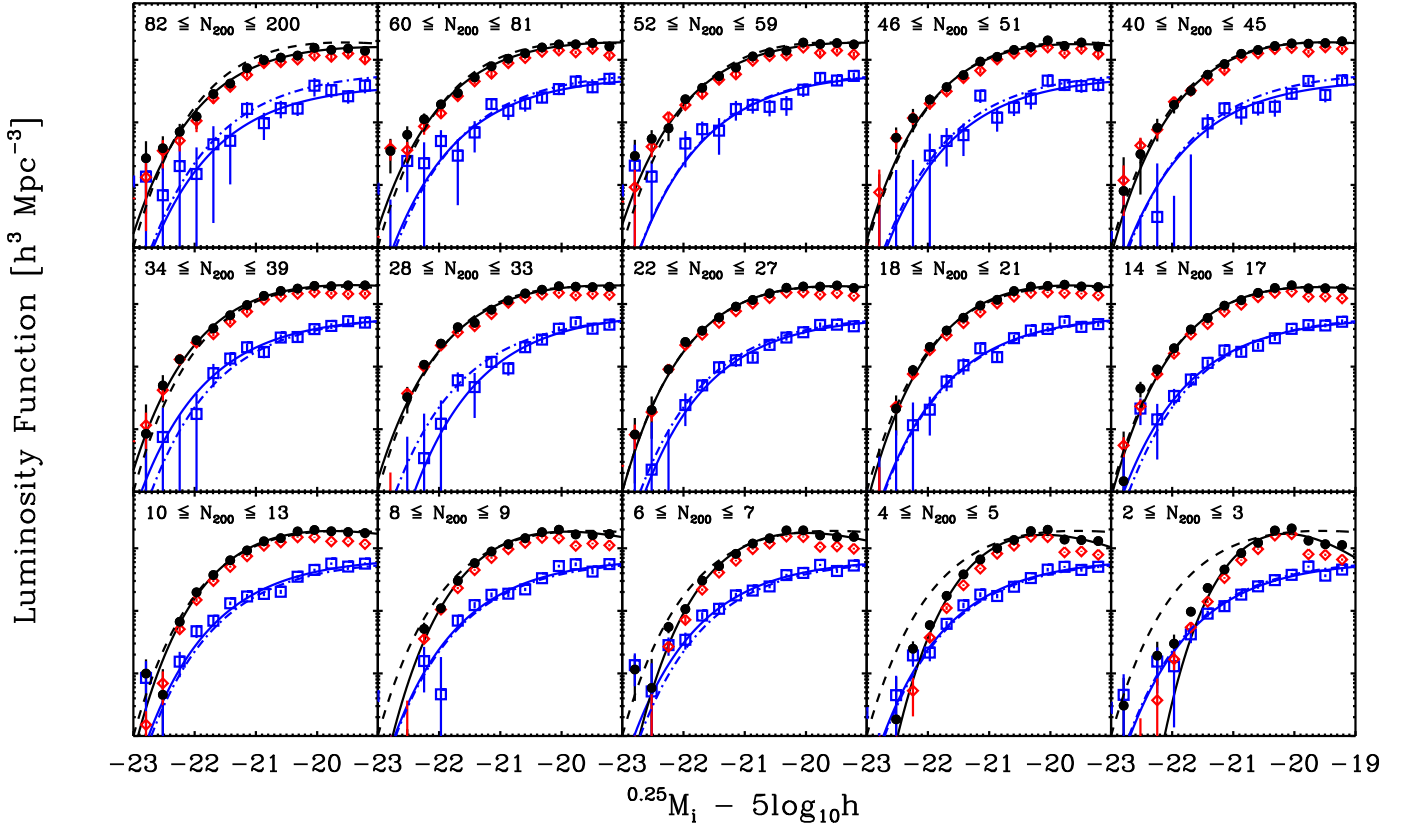} \figcaption[lfr200_colsplit.eps]{The
    luminosity function of all, red, and blue satellite galaxies in
    clusters (within \rtwo), as a function of
    cluster richness. Each panel represents a bin in cluster richness as given in the legend.
    Black filled circles show the LF for all satellites, with the best-fit
    Schechter function to those data shown as the solid black
    line. The red and blue galaxy samples are shown by red diamonds and blue
    squares respectively.  While the majority of satellites are red,
    there is a population of blue galaxies in clusters of all
    richnesses. Also shown is the Schechter function with fixed
    faint-end slope ($\alpha_{blue} = 1.0$) fit to the blue
    galaxy distribution (solid blue line). For reference, the
    best-fitting Schechter function for the mean LF of all satellite
    galaxies in clusters where the catalog is known to be complete and
    pure (\Ntwo\ $\ge 10$) is reproduced in each panel for all
    satellites (dashed black line) and blue satellites (blue
    dot--dashed line). The Schechter fits to the red galaxies are
    omitted for clarity. \label{fig:clf_colsplit}}
\end{figure*}

\section{Results}\label{sec:results}

Our primary results comprise the luminosity and color distributions
for cluster galaxies as a function of cluster richness,
cluster-centric distance, and redshift.  We examine BCG and satellite
galaxies separately, motivated by the observational and theoretical
reasons previously discussed. As an example of BCG-satellite
differences, Figure \ref{fig:LFwithBCG} shows the luminosity function
for all galaxies within \rtwo\ of cluster centers for four different
bins of cluster richness.  The luminosity functions of the satellites
(filled circles) and BCGs (open diamonds) are shown separately; the
total LF is given by the black line (with error bars omitted for
clarity).  Examination of the $\chi^2/d.o.f.$ reveals that the
satellite LF is statistically well described by a Schechter function,
but this behavior is not the case when the BCG is included.  The
BCG-only LF is well-fit by a single Gaussian.  In lower richness
clusters, the BCGs dominate the total light but are systematically
fainter and have larger luminosity variance than BCGs in larger
systems. These differences illustrate one way in which BCGs tend to be
distinct from satellites.

We begin by presenting results describing the satellite population,
then turn to the BCGs.  In \S\ \ref{sec:sats}, we examine the
luminosities and colors of satellites. Specifically, we measure the
luminosity function of satellies within \rtwo\ as a function of
cluster richness; this measurement is closely related to the
conditional luminosity function that is parametrized as a function of
mass. We also investigate the CLF of red and blue satellites
separately, and show the radial trend of the LF for all, red, and blue
satellites. In addition, we measure the red fraction of satellites and
investigate its dependence on several cluster properties. In \S
\ref{sec:BCGs} we explore the BCG population and quantify the
relationships of BCG luminosity to total cluster mass and to satellite
total and characteristic luminosity.

\subsection{Satellite Cluster Galaxies}\label{sec:sats}

In this section we measure the conditional luminosity function and its
dependence on galaxy color and cluster-centric distance. We also study how the
red fraction \fred\ depends on redshift, cluster mass, cluster-centric
distance, and galaxy luminosity.

To split the galaxies into red and blue subsamples, we make a cut in
color--luminosity space:
\begin{equation}\label{eq:colsplit}
^{0.25}(g-r) = 0.17 \log_{10}(^{0.25}L_i/(L_{\odot} h^2)) - 0.6.
\end{equation}
Because of the strong bimodality in the population, our results do not
depend sensitively on the exact placement of the red--blue
boundary. Figure \ref{fig:cmsplit} shows the bivariate distribution of
color and luminosity for four example richness bins and two radial
ranges: $3 \le $ \Ntwo\ $\le 5$; $10 \le $ \Ntwo\ $\le 15$; $30 \le $
\Ntwo\ $\le 40$; $80 \le $ \Ntwo\ $\le 200$ and $0.05 \le r/$\rtwo\
$\le 0.3$ (left set of 4 panels); $3.0 \le r/$\rtwo\ $\le 5.0$ (right
set of four panels) respectively.  The red sequence exists at all
richnesses. Even at large radii, there is an excess over random of
galaxies associated with clusters, and many of them are blue.  The
dashed line in the figure shows our adopted cut between red and blue
galaxies.

\begin{figure}
  \plotone{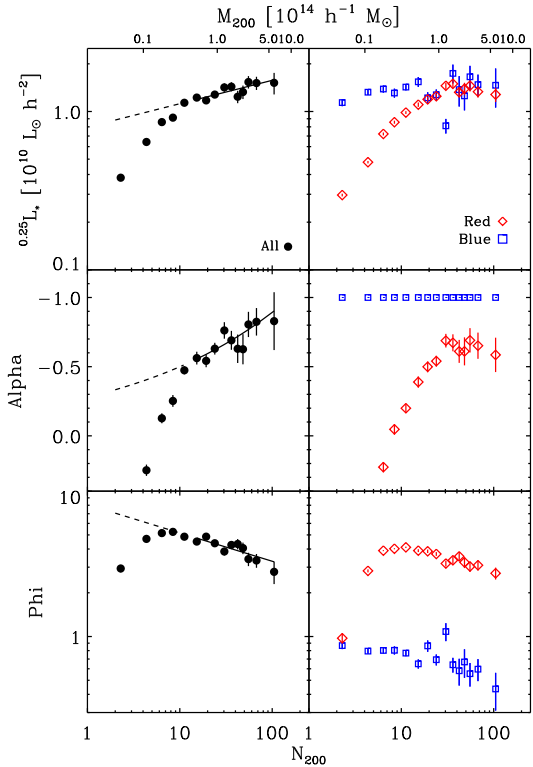}
  \figcaption[lstar_ngals.eps]{Parameters of the best-fitting
    Schechter functions from Figure \ref{fig:clf_colsplit} for all
    satellites ({\bf left}) and for red and blue satellites ({\bf
      right}) as a function of cluster richness. Symbols are the same
    as in Figure \ref{fig:clf_colsplit}. {\bf Top:} Characteristic
    galaxy luminosity, \lstar; {\bf middle:} faint end slope,
    $\alpha$, which is fixed to 1.0 for blue satellites; {\bf bottom:}
    normalization of the LF, $\phi_*$. The best-fitting power law
    describing the Schechter parameters as a function of richness for
    all satellites is shown as a solid line where the fit was
    performed, and as a dashed line extending the trend to lower
    richness. The power law parameters may be found in Table
    \ref{tab:schparamfit}.
  \label{fig:lstarngals}}
\end{figure}

\subsubsection{Conditional Luminosity Functions}\label{sec:CLF}

We start by investigating the dependence of total, as well as red and
blue split, luminosity functions on cluster richness and
cluster-centric distance. We use bins defined by sharp cuts in
richness; due to scatter in the mass--observable relationship, each
bin therefore contains clusters from a non-sharply defined mass range.
Note that these observationally-defined conditional luminosity
functions, which depend on both richness and cluster-centric distance,
are related to but differ somewhat in definition from the traditional
definition of the CLF as a function of cluster mass where the binning
is done in specific mass ranges \citep[as in, \eg,][]{Yang03}.

Figure \ref{fig:clf_colsplit} shows the luminosity function of
satellite galaxies within \rtwo\ for different bins of cluster
richness. These LFs are shown for all satellites (filled
circles) and for the blue and red subsamples separately (open squares
and open diamonds respectively).  The shape of the overall satellite
LF is only weakly dependent on \nvir\ for the top two rows,
representing higher \nvir\ clusters. These are the systems that have
been well studied and are known to be quite complete and pure.  For
\Ntwo\ $< 10$, sub- and super-\Lstar\ satellites are both found with
lower density than in more massive systems, but the incidence of
\Lstar\ galaxies is about the same as in the higher-mass objects. With
the split into red and blue subsamples, we see that change in the
faint end of the overall LF is driven by the changing ratio of red and
blue galaxies. We investigate the changing fraction of red galaxies as
a function of galaxy luminosity and cluster mass further in \S\
\ref{sec:BF}.

\begin{figure*}
  \epsscale{0.85} \plotone{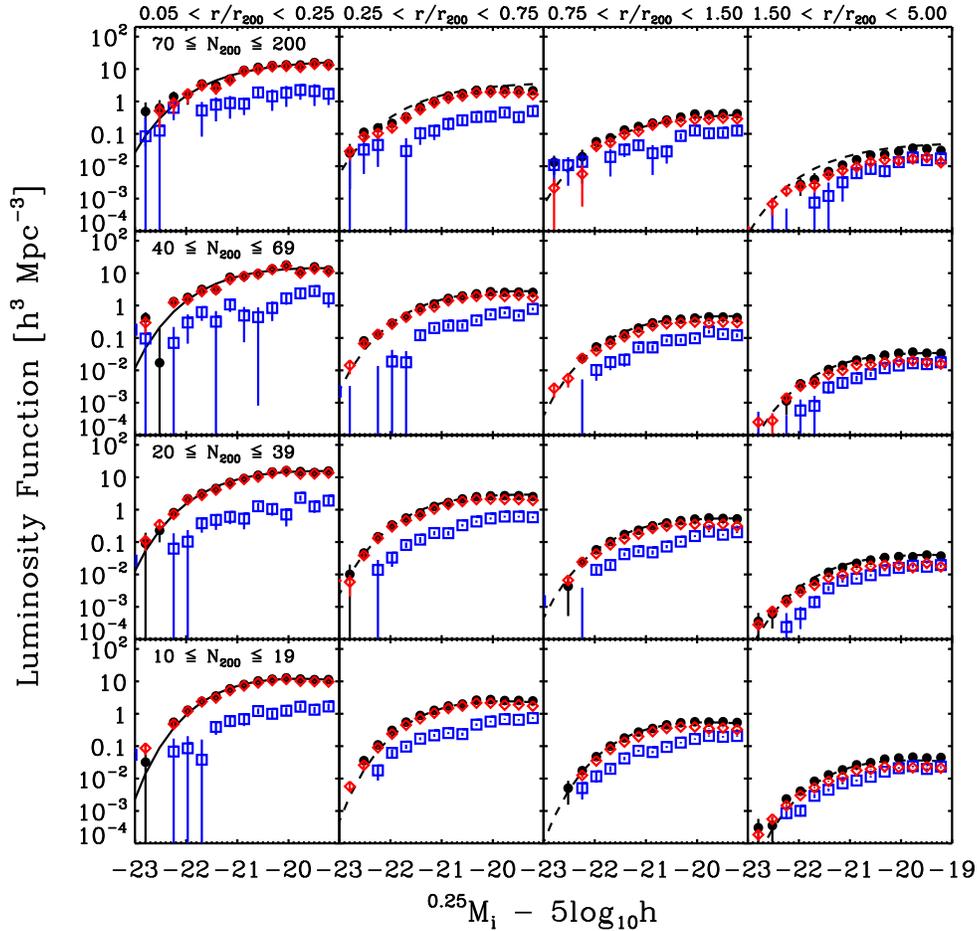}
  \figcaption[lf_radial_ngals.eps]{The luminosity function of all
    (black filled circles), red (red open diamonds), and blue (blue
    open squares) satellites as a function of distance from cluster
    center (increasing across rows) and of cluster richness
    (decreasing down columns). The best-fit Schechter function to all
    satellites in the innermost radial bin of each richness range
    (solid black lines in left-most column) is rescaled by $\phi_*$
    for the other radial bins (dashed lines). The excellent match
    between this scaled LF and the data illustrates that the LF
    changes normalization, but not shape, as a function of radius for
    these magnitudes. However, the fraction of red galaxies steadily
    increases with increasing distance from the center, most
    noticeably for sub-\lstar\ galaxies. \label{fig:lfrad}}
\end{figure*}

We fit the above LFs with \citet{Schechter76} functions, 
\begin{eqnarray}
\lefteqn{\phi(M)dM=}\nonumber\\
&& 0.4{\rm ln}(10)\phi_{*}10^{-0.4(M-M_{*})(\alpha+1)}e^{-10^{-0.4(M-M_{*})}}dM
\end{eqnarray}
using the Levenburg-Marquardt $\chi^2$ minimization procedure.  The
solid black lines in Figure \ref{fig:clf_colsplit} show the
best-fitting Schechter function for the total satellite LF in each
richness bin. The satellite LF is well fit by a Schechter function
except for in the very lowest richness bin.  For reference, the
best-fit Schechter function to the mean satellite LF for clusters with
\Ntwo\ $\ge 10$ is repeated in all panels as a dashed line.  We also
fit a Schechter function to the LF of the blue and red galaxy
populations.  For blue satellites, in the cases where the fit is well
constrained, we find the faint end slope to be $\alpha_{blue} = 1.0$.
In some cases the blue data do not permit a strong constraint on the
faint-end slope. We therefore fit the blue LFs with a faint-end slope
fixed to be 1.0.  These Schechter functions for blue satellites is
shown in each panel as the solid blue line, and for comparison, the
best fit for the mean blue satellite LF of clusters with \Ntwo\ $\ge
10$ is repeated in all panels as a dot--dashed line. The fits for the
red satellites are omitted for the sake of clarity.

\begin{deluxetable}{lcccc}
\tabletypesize{\small}
\tablecaption{Power Law Fits to Schechter Parameters \label{tab:schparamfit}} 
\tablewidth{0pt}
\tablehead{
  \colhead{Parameter} &
  \colhead{Units} &
  \colhead{Vs.} &
  \colhead{$Normalization$} &
  \colhead{$Index$}
}
\
\startdata
$^{0.25}$\Lstar ..... & $10^{10}L_\odot$   & \Ntwo    & 0.8 $\pm$ 0.1    & 0.15 $\pm$ 0.04 \\
                & $10^{10}L_\odot$   & $M_{14}$ & 1.29 $\pm$ 0.02  & 0.12 $\pm$ 0.03 \\
$\alpha$ ............ & & \Ntwo    & -0.28 $\pm$ 0.06 & 0.25 $\pm$ 0.07 \\
         & & $M_{14}$ & -0.63 $\pm$ 0.02 & 0.20 $\pm$ 0.05 \\
$\phi_*$ .......... & $h^3$ Mpc$^{-3}$ & \Ntwo    & 8 $\pm$ 1        & -0.20 $\pm$ 0.04 \\
         & $h^3$ Mpc$^{-3}$ & $M_{14}$ & 4.3 $\pm$ 0.1    & -0.16 $\pm$ 0.03 \\

\enddata
\tablecomments{$M_{14} \equiv M_{200}/(10^{14} M_\odot)$}
\end{deluxetable}

The parameters of these best-fitting Schechter functions are shown in Figure
\ref{fig:lstarngals} as a function of \nvir.  For systems with \Ntwo\ $\geq$
10, $^{0.25}$\lstar\ for all satellites is a weak function of mass, and is
roughly $1.4 \times 10^{10} h^{-2} L_{\odot}$. For blue galaxies in clusters of any
richness, $^{0.25}$\lstar\ is consistent with $1.4 \times 10^{10} h^{-2} L_{\odot}$;
red galaxies have \lstar\ consistent with this value only for \Ntwo\ $\gtrsim$
30 ($10^{14}$\Msun), while lower mass systems have red satellites with a
fainter characteristic luminosity. The faint end slope, $\alpha$, scales with
mass for all and red satellites, with a steeper faint end slope for more
massive clusters. For blue galaxies, we have fixed the value of $\alpha$, but by
inspection of Figure \ref{fig:clf_colsplit} it is seen that $\alpha_{blue}$
changes little over the whole richness range of the catalog.  The normalization
of the LF, $\phi_*$, decreases with mass for all, red, and blue galaxies. For
each of \lstar, $\alpha$ and $\phi_*$, we fit the trend of the parameter as a
function of richness for \Ntwo\ $\ge 10$.  The fits are shown on the Figure,
with a solid line where the fit was performed and a dashed line extending the
relationship to lower richness. The parameters of these fits are listed in Table
\ref{tab:schparamfit}.

These CLF results show the weak dependence of the LF parameters on
cluster mass, and are in reasonable agreement with other recent
modeling and observational results.  The relationship between
characteristic satellite luminosity and cluster richness was recently
discussed by \citet{Skibba07}.  They explored the predictions of
\citet{Skibba06} that, for halos with mass greater than
$10^{12}$\Msun, their Halo Occupation Distribution (HOD) models predict that the mean luminosity of
non-central galaxies should be nearly independent of mass and that the
shape of the LF is approximately independent of mass.
\citet{Skibba07} showed that the mean non-central galaxy luminosity is
indeed only very weakly dependent on cluster richness (as quantified
by the number of galaxies with $M_r < -19.9$) in the group catalogs of
\citet{Yang05} and \citet{Berlind06}.  The catalogs used in
\citet{Skibba07} just barely overlap in mass with the lower mass end
of our \Ntwo\ $\ge 10$ sample. Our findings for \Ntwo\ $\ge 10$ are in
agreement with the group-scale results, and show that the HOD
prediction continues to hold for greater mass systems as well.  The
results of CLF modeling discussed in \citet{Cooraydivided} and
\citet{Cooray06}, based on constraints from the two-point correlation
function, also qualitative agree with this the behavior of \Lstar\ as
a function of cluster mass.

\begin{figure*}
  \plotone{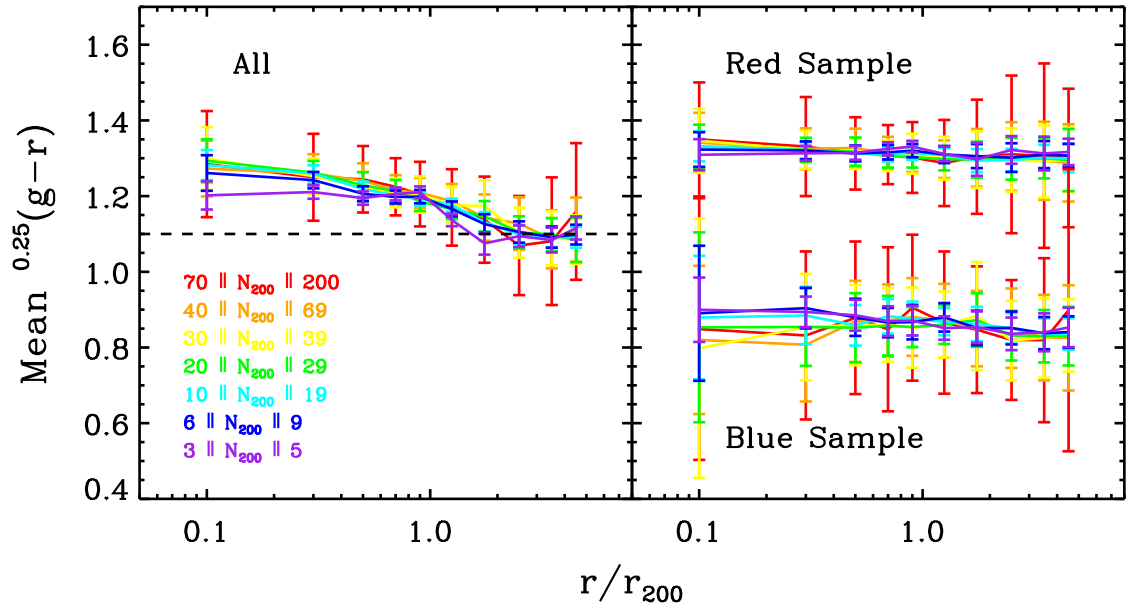} \figcaption[col_radial_incolor.eps]
	  {Mean \gmr\ color as a function of \rad\ changes due to the
	  changing red and blue fractions.  {\bf Left:} all
	  satellites; {\bf right:} the sample split into red and blue
	  galaxies.  The mean color overall changes, but as the mean
	  color of the red and blue samples does not change, the
	  overall shift must be due to the changing numbers of red and
	  blue galaxies.  The estimated global field value is shown by
	  the dashed line. \label{fig:col_radial_red_blue}}
\end{figure*}

The dependence of the faint end slope on \nvir\ has also been
discussed elsewhere; \citet{Cooraydivided,Cooray06} also found a
steeper faint end slope for more massive systems, and a changing faint
end slope was incorporated into the CLF models of \citet{vdB05} in
a similar way to fit the observed trends in 2dFGRS data. The trend for
the faint end falling off more rapidly in less massive systems is
qualitatively similar to the trend seen in RCS-1 data
\citep{Gilbank07}, although more work is required to make a detailed
comparison with this higher-redshift, and very differently selected,
sample.  We note that the magnitude limit of our data is a relatively
bright \Mi\ $< -19$, so we cannot comment on the faint end turn-up
seen in deeper studies \citep{Yagi02,PopessoLF06,Barkhouse07}. Our
results for all, red, and blue satellites are in good agreement with
these deeper results in the magnitude range where we overlap.

The somewhat counter-intuitive trend of decreasing $\phi_*$ in more
massive systems reflects the correlation between (M/L)$_{200}$, the
mass-to-light ratio measured at \rtwo, and cluster mass. The
(M/L)$_{200}$--\Mtwo\ relationship is both expected theoretically in
some models \citep[\eg,][and references therein]{Tinker05,vdB07} and
directly observed \citep{Sheldon07b}. By definition, the mass density
within \rtwo\ is the same for clusters of all masses. Since
(M/L)$_{200}$ increases as a function of cluster mass, the
luminosity density must be decreasing with mass.  As the typical
galaxy luminosity \Lstar\ increases with mass, the mean number density
of galaxies within \rtwo\ must decrease with mass.

In \citet{Hansen05} we explored the LFs of galaxies in clusters as a
function of richness, and it was not clear that all satellite LFs
within \rtwo\ were well described by a Schechter function, especially
for the low-richness systems. However, the cluster finder used in that
previous work used a fixed 1 Mpc aperture for estimating richness,
resulting in much greater scatter in the mass--richness relation than
in the present version of the cluster-finding algorithm, thereby
mixing together clusters of much different mass into the same bin of
richness. In addition, as we will show, the LF normalization changes
dramatically as a function of $r/$\rtwo, so a fixed aperture will mix
populations from different fractional \rad\ sections of clusters and
thus further confuse the analysis.  While any catalog will have noise
and impurities, the current cluster catalog, which uses and improved
richness measure, should be less affected by this issue.

In addition to the LF within \rtwo, we also investigate the radial
dependence of the luminosity function for all galaxies and for red and
blue galaxies separately in several bins of cluster richness. These
results are shown in Figure \ref{fig:lfrad}.  Cluster richness
decreases downward in the array of plots, and the radius range
increases to the right.  There are still galaxies correlated with the
clusters centers at separations much larger than \rtwo.  While these
galaxies are likely not bound to the clusters, they are statistically
correlated with clusters and the amplitude of the luminosity function
reflects the amplitude of the cluster-galaxy cross-correlation
function.

The normalization of the LF decreases with cluster-centric distance,
but the shape of the total satellite LF remains essentially unchanged
for the magnitude range considered here. To illustrate this point, we
have re-drawn the best-fit Schechter function for all satellites in
the innermost radial bin (solid lines in left hand column) in the
panels for other radial bins, but rescaled by $\phi_*$ (dashed lines
in other columns). This rescaled LF is an excellent match to the
measurements. For reference, we find that in the $0.25 < r/r_{200} <
0.75$, $0.75 < r/r_{200} < 1.5$ and $1.5 < r/r_{200} < 5$ radial bins
respectively, $\phi_*$ is typically 20\%, 3\%, and 0.3\% of $\phi_*$
in the innermost bin.

Although the relative mix of luminosities is unchanged with radius,
there is a strong change in the relative mix of colors.  Red
galaxies are much more dominant in the inner regions, and the
decreasing contribution of red galaxies to the total LF at large
radius is driven by changes in the faint end slope of the red LF. The
shape of the blue galaxy luminosity function changes relatively little
with radius, and at all radii the blue LFs show
proportionately more faint galaxies than do the red or total LFs. The
relative constancy of the bright end of the LF of all satellites as a
function of radius is in agreement with the recent detailed
examination of 57 low-redshift Abell clusters by \cite{Barkhouse07},
as is the finding that the LFs of red galaxies change more
dramatically with radius than do the blue LFs.  

These trends are discussed further in the context of the red fraction measurements below.

\subsubsection{The Red Fraction \& Other Color Statistics}\label{sec:BF}

From examination of the LFs, it is clear that the red fraction changes with a
variety of parameters. In this section we examine the dependence of the red
fraction on cluster richness, cluster redshift, cluster-centric distance, and
satellite galaxy luminosity.

To assess whether changes in the galaxy population are primarily
related to changes in the red fraction, we examine the typical color
of satellites as a function of distance from the BCG.  Figure
\ref{fig:col_radial_red_blue} shows the (number weighted) mean \gmr\
color for satellite galaxies with \Mi\ $< -19$, in bins of
cluster-centric distance out to 5$\times$\rtwo, and for several bins
in cluster richness. The left panel shows the trend for all galaxies,
while the right-hand panel shows the mean color for the red and blue
samples of galaxies separately.  The dashed line shows the estimated
mean color of the universal average population.  The overall mean
satellite color is redder in the inner regions and trends bluer as a
function of radius until $r\sim 2 \times$\rtwo, beyond which the mean
color is consistent with the universal average value. The mean color
is insensitive to cluster richness at any radius, with the exception
of the very lowest richness bin in which the galaxies are typically
bluer than in richer clusters.  However, the right panel shows that
the red and blue samples remain the same color regardless of
cluster-centric distance or cluster richness, which implies that the
overall shift in color is simply a change in the relative number of
red and blue galaxies.  This observation suggests either that there is
some process that distributes red and blue galaxies differently with
radius, or that the transition between the two populations happens
fairly rapidly.  Regardless of the mechanism, the color shift appears
to be characterized well by the changing number of red galaxies, and
we subsequently consider the red fraction, \fred, as the primary
statistic for quantifying the change in galaxy population.

\begin{figure}
  \epsscale{1.2} \plotone{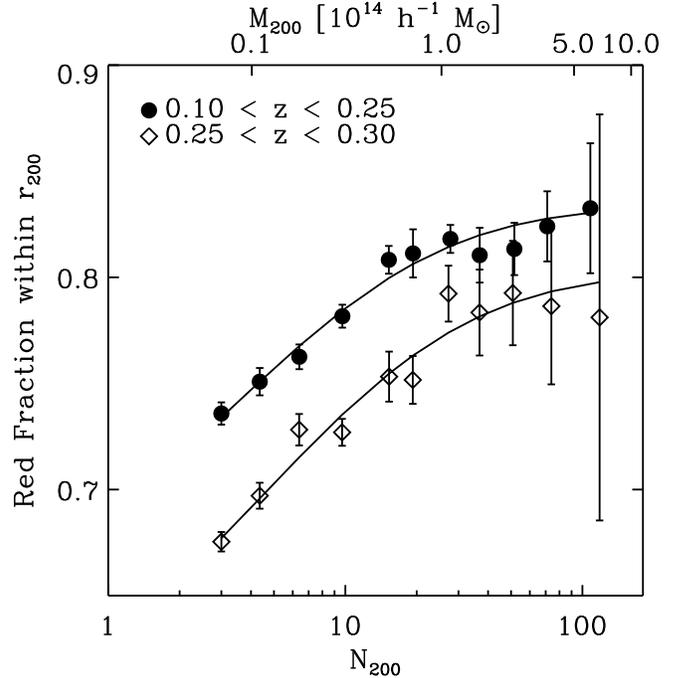}
  \figcaption[rfrac_r200_zbins.eps]{Red fraction within \rtwo\ (\Mi\
  $< -19$) as a function of richness for two redshift bins: $0.1$ \lae\ $z <
  0.25$ (black filled circles, median $z = 0.2$) and $0.25$ \lae\ $z$ \lae\ $0.3$
  (brown open diamonds, median $z = 0.28$). The solid lines are the
  richness and redshift dependent model of Equation \ref{eq:fredfit}. \label{fig:bfrac_r200}}
\end{figure}

We first measure the satellite red fraction within \rtwo\ as a
function of cluster richness and redshift.  Figure
\ref{fig:bfrac_r200} shows the fraction of red satellites brighter than \Mi\ $=
-19$ and within \rtwo\ as a function of cluster richness.  The clusters
are split into two bins of redshift: $0.1 \le z < 0.25$ (filled
circles, median $z = 0.2$) and $0.25 \le z \le 0.3$ (open diamonds,
median $z = 0.28$).  The red fraction increases significantly as a
function of cluster mass in both redshift slices, but for lower
redshift systems \fred\ is systematically higher at all richnesses by $\sim$5\%. There
is a hint that \fred\ flattens above $10^{14}$\Msun. To
parametrize the trend of red satellite fraction as a function of
cluster richness and redshift, we adopt the functional form
\begin{equation}
f_R = g(z)\,{\rm erf}[\log_{10}(N_{200}) - \log_{10}(h(z))] + 0.69 \label{eq:fredfit}.
\end{equation}
We find that with
\begin{eqnarray}
g(z) = 0.206 -0.371 z\\
h(z) = -3.6 + 25.8 z
\end{eqnarray}
this form provides a reasonable description of the data; this relationship is
shown as the solid lines in Figure \ref{fig:bfrac_r200}. For \Ntwo\ $<
10$, the catalog may be less than 90\% complete and pure, and it is
possible that as-yet-unquantified selection effects may influence the
trend in this regime.

Here, we have made measurements within the aperture \rtwo\ measured
for the full cluster sample with the lensing data, relative to the
critial density at $z = 0.25$. However, as the critical density scales
with redshift, one can question whether we have biased our result by
not using the \rtwo\ taken relative to the critical density
appropriate for the mean redshifts of the two subsamples. While the
lensing analysis has not been done separately for the two redshift
ranges, we can investigate the dependence of \rtwo\ on redshift using
the galaxy density estimate. We find that using the $z$-appropriate
threshhold for the high and low redshift samples results in \rtwo\
values that are 8\% and $<$ 1\% different, respectively, from the
value found for the whole sample. Using an adjusted aperture size, we
measure red fractions that are 0.5\% different for the higher redshift
sample. The change in the low redshift sample is even smaller. This
systematic offset is clearly not sufficient to explain the observed
5\% difference in red fraction between the the two redshift bins.

\begin{figure*}
  \plotone{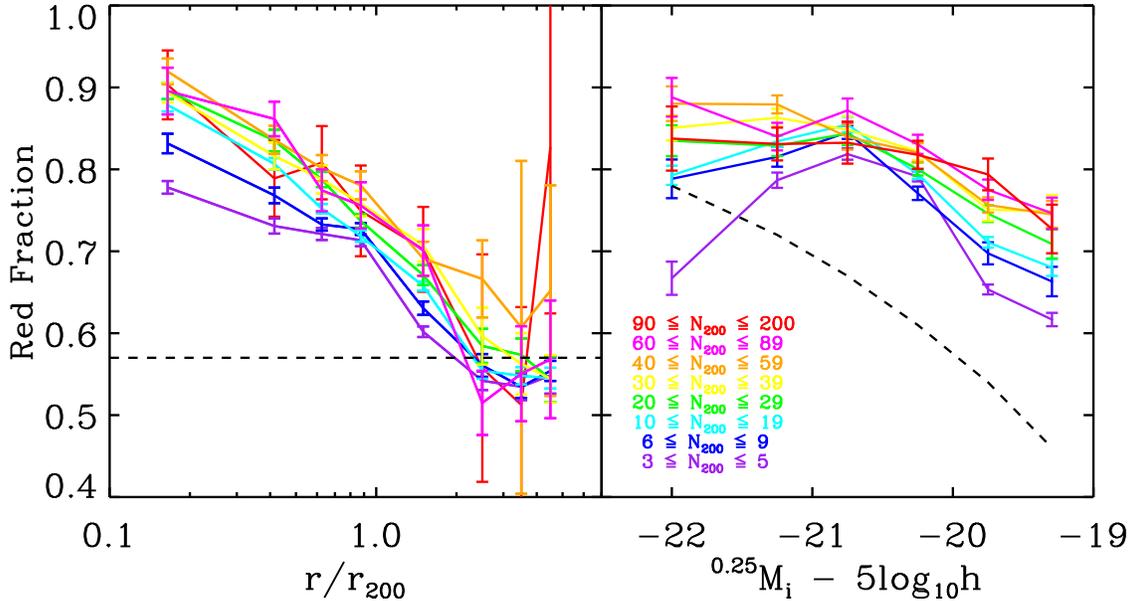} \figcaption[rfrac_rad_mag_incolor.eps]{Dependence
    of red fraction on cluster-centric distance and galaxy
    luminosity. {\bf Left:} Red fraction (for galaxies with \Mi\ $< -19$) as a function
    of \rad\ for different richness bins.  The estimated global field
    value is shown by the dashed line.  {\bf Right:} Red fraction
    (within \rtwo) as a function of absolute magnitude for different
    richness bins.  The estimated global field value for each
    luminosity is shown by the dashed line.
\label{fig:bfrac_rad_rich_mag}}

\end{figure*}

As seen from the behavior of the red and blue LFs, the red fraction has additional dependence on
cluster-centric distance and satellite galaxy luminosity.  In Figure
\ref{fig:bfrac_rad_rich_mag} we examine both the radial and luminosity
dependence of the red fraction. The left panel of Figure
\ref{fig:bfrac_rad_rich_mag} shows \fred $(r/$\rtwo$)$ for satelllies with \Mi\
$< -19$ for clusters in several bins in richness. Within $\sim 2\times$\rtwo,
the red fraction decreases with radius.  Beyond $2\times$\rtwo, \fred\
flattens asymptotically, approaching the cosmological average (dashed line). The
red fraction is essentially independent of cluster richness for \Ntwo\ $\ge 10$
systems, but is correlated with richness for \Ntwo\ $< 10$.

The right panel of the Figure shows \fred\ for satellites within \rtwo\ as a
function of absolute magnitude for multiple richness bins. For satellites
brighter than \lstar, the red fraction is relatively independent of luminosity
except for the lowest richness systems.  For sub-\lstar\ galaxies, the red
fraction steadily decreases toward fainter magnitudes, and the trend is
stronger for lower \nvir\ systems.  Except for the brightest galaxies in the
very lowest-richness bin, \fred\ is always greater than the universal average
value (dashed line).

Broadly speaking, the fraction of red satellite galaxies increases
with cluster mass, with galaxy luminosity, with time, and for
decreasing cluster-centric distance.  The dependence on cluster
richness is coupled more strongly to satellite luminosity than to
cluster-centric distance.  The previous LF results showed that the
faint ends of the red and blue satellites' LFs change as a function of
cluster-centric distance, although the bright ends remain essentially
unchanged. In light of this observation, we conclude that it is the
steadily changing mix of sub-\Lstar\ galaxies as a function of
distance from the cluster center that causes the trend of decreasing
\fred\ with cluster-centric distance.

These results are in general agreement with previous work, despite
differences in sample definitions for both cluster selection and
sample splitting. Several studies
\citep{Margoniner01,Goto03,Weinmann06a,Poggianti06,Martinez06,Weinmann06b,Gerke07,Desai07,BlantonBerlind07}
have reported that the galaxy type fraction is a function of cluster
mass.  However, others \citep{Balogh04b,depropris04, Tanaka04} found
that the type fraction does not depend on cluster velocity dispersion,
although \citet{Weinmann06a} argues that this result may be due to the
large scatter in relationship between velocity dispersion and mass.
Furthermore, the flattening and ``saturating'' of the red fraction for
the most massive systems appears to be a natural product of modeling
\citep{Berlind03, Zheng05}.  Some authors
\citep{depropris04,Weinmann06a,BlantonBerlind07} have also found
qualitatively similar behavior for the type fraction as a function of
radius; however, this trend is very weak in both the \citet{Wilman05}
and \citet{Gerke07} data, but there are large uncertainties in both of
those data sets that could mask a correlation. The luminosity
dependence of the galaxy type fraction is also similar to previous
results \citep{depropris04, Wilman05, Weinmann06a, Martinez06,
  Gerke07}.

The redshift dependence of the galaxy type fraction has been discussed
by many authors since \citet{ButcherOemler}.  Overall, we find that
the fraction of red galaxies increases by $\sim$5\% during the 0.8
Gyr between our higher (median $z=0.28$) and lower (median $z=0.2$)
redshift samples over the full mass range. This result is in quite
good agreement with the results of the eponymous work, where an
increase in the blue fraction by 0.25 in clusters to $z = 0.5$ was
observed \citep{ButcherOemler}. We also find agreement with
\citet{Goto03}, who used a smaller catalog of clusters from the SDSS
EDR. With the larger \maxbcg\ sample, however, we can see that this
shift in red fraction happens in a similar manner for clusters of all
richnesses. Direct comparison with higher redshift samples, such as
the blue fraction of the DEEP2 groups measured by \citet{Gerke07}, is
not straightforward because of the very different sample selection
criteria used, but in future work we will undertake this comparison.

\subsection{Luminosity of Brightest Cluster Galaxies}\label{sec:BCGs}

We now examine the Brightest Cluster Galaxies, focusing on their
luminosities and how those luminosities compare to the total cluster
luminosity and the characteristic satellite luminosity.

\begin{figure}
    \epsscale{1.2}
    \plotone{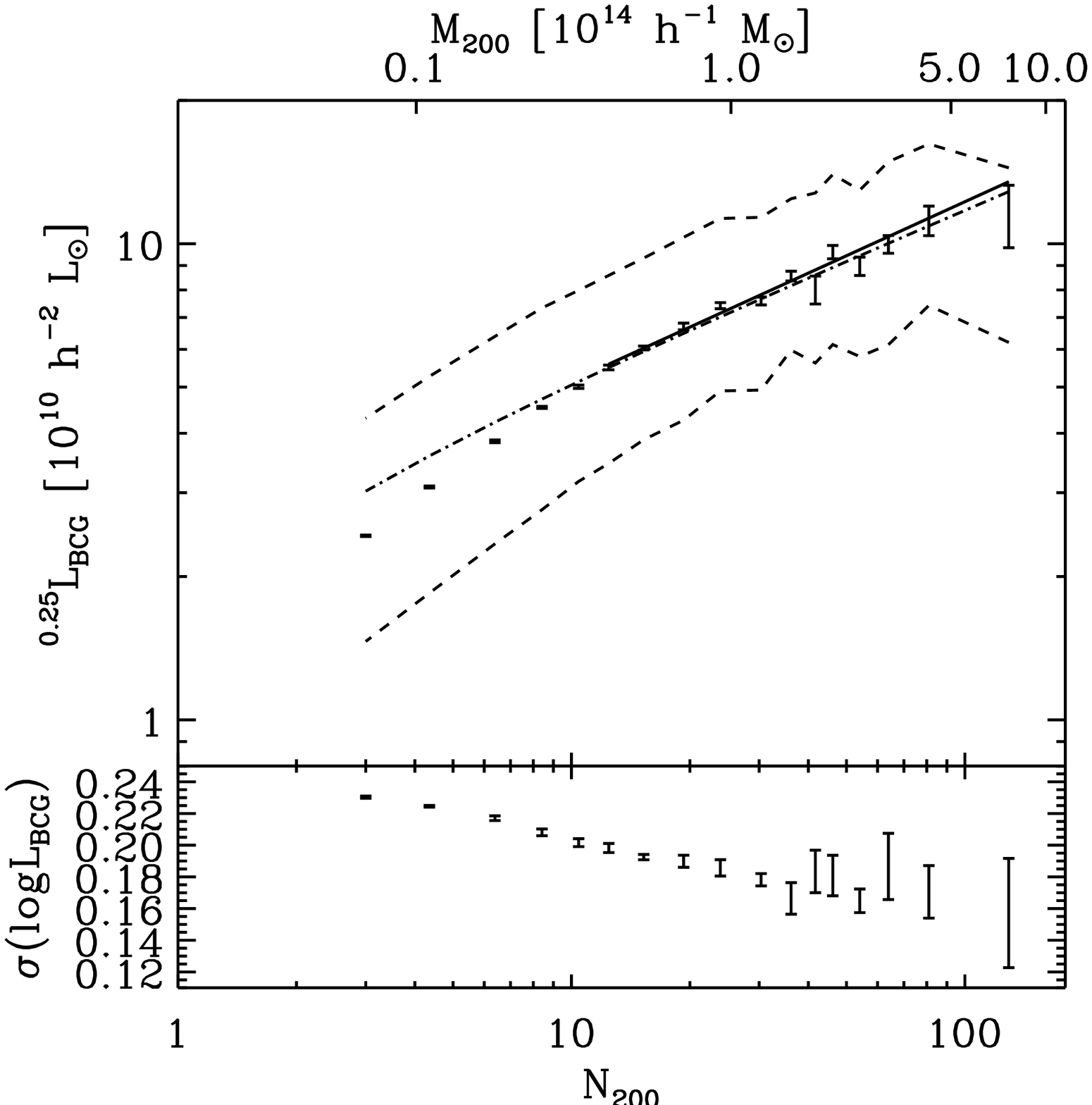} 
  \figcaption[bcglum_ngals.eps]{{\bf Top:} Median BCG $^{0.25}i$-band
  luminosity, \lbcg, as a function of richness. Error bars show the statistical
  uncertainty on the luminosity; the dashed lines show the 68\% scatter in
  \lbcg\ in each richness bin. The solid line shows the best-fit power law,
  \Lbcg\ $\sim M_{200}^{0.3}$, fit for \Ntwo\ $\ge 10$. The dot--dashed line
  shows the best-fitting \citet{VO06} model. {\bf Bottom:} The width of the
  68\% scatter of \lbcg\ as a function of richness. For massive systems this
  width is $\sim$ 0.17.\label{fig:lbcg_ngals}}
\end{figure}

The top panel of Figure \ref{fig:lbcg_ngals} shows the median luminosity of BCGs,
\Lbcg, as a function of cluster richness.  Systems with a greater number of
satellites tend to host brighter BCGs. The error bars on the data points
indicate the statistical uncertainty on the median BCG luminosity in each bin,
determined from jackknife resampling; the dashed lines show the region within
which 68\% of the BCG luminosities lie. For \Ntwo\ $ \ge 10$, we parametrize
the \Lbcg-richness trend with a power law, and find that the $i$-band light from BCGs scales
with cluster richness as
\begin{equation}
^{0.25}L_{BCG} = (2.16 \pm 0.08) \times 10^{10} h^{-2} L_{\odot}\, N_{200}^{0.38 \pm 0.01}. 
\end{equation}
This fit is shown as a solid line Figure \ref{fig:lbcg_ngals}. Using our
adopted mass--observable scaling, this relationship is 
\begin{equation}
^{0.25}L_{BCG} = (6 \pm 2) \times 10^{10} h^{-2} L_{\odot} \left(\frac{M_{200}}{M_{14}}\right)^{0.30 \pm 0.01}, 
\end{equation}
where we have defined $M_{14} \equiv 10^{14}$\Msun.
A scaling suggested by \citet{VO06} for the $L_{BCG}$-mass relation is
\begin{equation}
\label{eq:Lcfit}
\langle L_c \rangle = L_0 \frac{(M/M_c)^a}{[1+(M/M_c)^{bk}]^{1/k}},
\end{equation}
with $M_c = 3.7 \times 10^9 h^{-1}M_{\odot}$, a = 29.78, b = 29.5, and
k = 0.0255. This functional form is derived by fitting the relationship obtained by
matching the galaxy luminosity function to the subhalo mass function,
and is not currently physically motivated.  Equation \ref{eq:Lcfit} also
provides an acceptable fit to our data if we adopt $L_0 = 4 \times 10^9 h^{-2}
L_{\odot}$, the mean galaxy luminosity in halos of $3.46 \times
10^{11}h^{-1}M_{\odot}$ in the $^{0.25}i$-band. This relationship is shown on the
Figure with the dot--dashed line. 

The trend of increasing central galaxy luminosity with cluster
richness has been noted in many previous observational studies
\citep[\eg,][]{Sandage73,Sandage76,Hoe80,Sch83}. New large cluster
samples with robust mass estimators have explored the scaling of BCG
luminosity with cluster mass.  Using a sample of 93 clusters with both
X-ray and $K$-band data, \citet{LinMohr04} found that BCG light scales
with mass as $L_{BCG,K-band} \sim M_{200}^{0.26}$ for clusters with
$M_{200} > 3 \times 10^{13} h^{-1} M_{\odot}$, with significant
scatter.  \citet{Yang05}, using groups found in the 2dFGRS, found that
in $b_J$-band, $\langle L_{cen} \rangle \sim M^{0.25}$ for halos of M
$> 10^{13}h^{-1}M_{\odot}$.  \citet{ZCZ07}, in their investigation of
the luminosity-dependent projected two-point correlation function of
DEEP2 and SDSS, also found that there is a correlation between halo
mass and central galaxy luminosity. Using the \citet{VO06} model, they
see that $L_0=2.8 \times 10^9 h^{-2} L_{\odot}$ provides a reaonable
fit for the SDSS $r$-band (with halo masses up to $3 \times 10^{13}
h^{-1} M_{\odot}$) and $L_0=4.3 \times 10^9 h^{-2} L_{\odot}$ is suitable for the
DEEP2 $B$-band data (with halo masses up to $4 \times 10^{12} h^{-1}
M_{\odot}$).  Using a different cluster catalog from the SDSS,
\citet{PopessoMLHOD} found $L_{BCG} \sim M_{200}^{0.33}$ for these 217
systems. Our results are in agreement with these findings within the
uncertainties, but the size of the \maxbcg\ catalog, its
well-understood selection function and its accurate mass estimator allow
us to probe the BCG luminosity distribution in further detail.

At all richnesses, the distribution of \Lbcg\ is well-described by a
Gaussian.  The mean value is dependent on cluster richness as
previously discussed. The width of the distribution, $\sigma_{log L}$,
is also a function of cluster richness.  The bottom panel of Figure
\ref{fig:lbcg_ngals} shows $\sigma_{log L}$ as a function of richness,
with error bars from jackknife resampling. There is an overall
negative correlation between $\sigma_{log L}$ and \Ntwo, although for
\Ntwo\ \gae 35, $\sigma_{log L}$ is roughly consistent with a constant
value of $\sim 0.17$. This value is somewhat higher than the
$\sigma_{log L} \sim 0.12$ found by \citet{ZCZ07}, but is consistent
within the uncertainties. Our measurements are made as a function of
cluster richness, and scatter in the mass--observable relation causes
clusters over some range of masses to be assigned to each
richness. Since \Lbcg\ depends on cluster mass, this scatter may result
in larger observed $\sigma_{log L}$ values than in the intrinsic
distribution of BCGs as a function of cluster mass. However, mass
mixing acts only to increase the observed value, so our measurements
represent at least a secure upper limit on the scatter in the
\Lbcg--\Mtwo\ relationship. Future work with simulations is required to
fully disentagle the intrinsic scatter from that introduced by the
mass proxy.

The \maxbcg\ cluster finder includes priors on BCG color and
luminosity (see \S\ \ref{sec:clustersamp}), and so we consider whether the
resulting BCG luminosity distribution is being artificially
constrained by these priors. In most cases, where BCG identification
is unambiguous, only the narrow color priors inform BCG selection, so
we do not expect a significant effect from the magnitude
prior. Detailed examination of the effect of these priors on the
selection function of \maxbcg\ is underway, but preliminary results
indicate that the incidence of rich clusters with ambiguous BCGs is
\lae\ 25\%, and does not have a significant effect on the scatter of
rich systems. Here, we see that the width of the distribution of
identified BCG $i$-band absolute magnitudes is $\sim$ 0.5mag wider
than the (mostly uninformative) magnitude prior, and that all BCGs are
easily brighter than the nominal 0.4\Lstar\ limit.  Thus, we take the
recovered distributions as representative of the cluster population,
but reserve a more detailed investigation for future work.

\begin{figure}
  \epsscale{1.2} \plotone{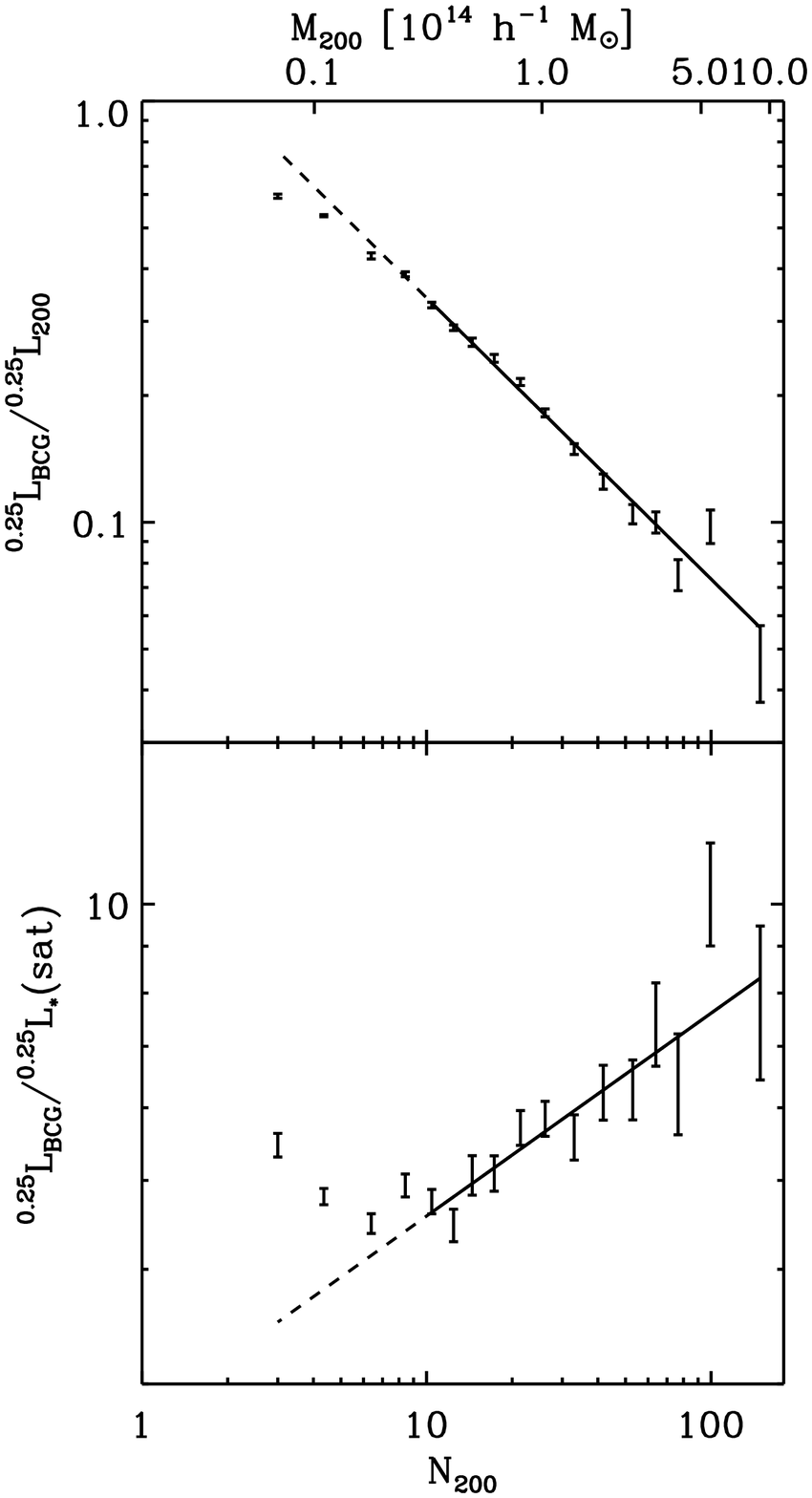}
  \figcaption[bcgcontrib_ngals.eps]{BCG luminosity compared to the
    luminosity of the rest of the cluster galaxies within \rtwo\ (\Mi\ $< -19$). {\bf
      Top:} Mean BCG luminosity fraction (\lbcg/\Ltwo) as a
    function of richness. {\bf Bottom:} The ratio of BCG luminosity to
    \lstar\ of the satellite galaxies, \Lbcg/\Lstarsat, as a
    function of richness. In each case the best-fitting power law is
    shown with a solid line where the fit was performed and a dashed
    line to extend the relation to lower
    richness. \label{fig:bcgcontrib_ngals}}
\end{figure}

In addition to examining the correlation between BCG luminosity and
cluster mass, we also measure the trend with mass of the ratios of BCG
luminosity to the total cluster luminosity, \Ltwo, and BCG
luminosity to the characteristic luminosity, \Lstarsat, of the
satellites. The top panel of Figure \ref{fig:bcgcontrib_ngals} shows
\Lbcg/\Ltwo\ as a function of cluster richness. As expected from
the example LFs of Figure \ref{fig:LFwithBCG}, \Lbcg/\Ltwo\
decreases with cluster richness. For the most massive clusters
($10^{15} h^{-1} M_{\odot}$), the BCGs supply only $\sim 5\%$ of the
cluster luminosity budget, but for intermediate $10^{14} h^{-1}
M_{\odot}$ systems the BCG makes up $\sim 20\%$ of the light.  For the
lowest richnesses the luminosity is completely dominated by the BCG.

Fitting a simple power law, we find that for clusters with \Ntwo\ $\ge 10$, the
BCG light fraction scales with cluster richness  as 
\begin{eqnarray}
    \frac{^{0.25}L_{BCG}}{^{0.25}L_{200}} = (1.58 \pm 0.06)\, N_{200}^{-(0.67 \pm 0.01)}
\end{eqnarray}
and with cluster mass as
\begin{eqnarray} \frac{^{0.25}L_{BCG}}{^{0.25}L_{200}} = (0.19 \pm 0.04) \left(\frac{M_{200}}{M_{14}}\right)^{-(0.53 \pm 0.01)}.  
\end{eqnarray}
This relationship is shown in the Figure with a solid line over the
\nvir\ range where the fit was performed, and as a dashed line to
extend the relation to lower richness.

The negative correlation between the BCG luminosity fraction and
cluster mass is in good agreement with other results. The $K$-band
data presented by \citet{LinMohr04} for their X-ray selected sample
expresses qualitatively the same trend. For a different catalog of
X-ray selected clusters, \citet{PopessoMLHOD} measured the observed
scaling between total optical luminosity and cluster mass and between
BCG luminosity and cluster mass; the combination of these
relationships yields a best fit scaling of \Lbcg/\Ltwo\ $\sim
M_{200} ^{-0.59}$, consistent with our results.  We are also
in reasonable agreement with the low redshift ($z=0.1$), $r$-band
model predictions of \citet{Iro07}, who expect \Lbcg/\Ltwo\
$\sim$ \Mtwo$^{1/2}$ for massive systems, and who also predict that
the normalization of this relationship may be sensitive to $\sigma_8$.

We also compare \Lbcg\ to the characteristic satellite luminosity
\Lstarsat. The value of \Lstarsat\ is derived from fits to Schechter
functions for satellites within \rtwo\ and brighter than \Mi\ $= -19$
as in \S\ \ref{sec:CLF}.  The bottom panel of Figure
\ref{fig:bcgcontrib_ngals} shows that the ratio \Lbcg/\Lstarsat\
depends strongly on \Ntwo. For clusters with \Ntwo\ $\ge 10$,
\Lbcg/\Lstarsat\ increases with \Ntwo, but this trend reverses for
very low richness systems. This behavior is not unexpected given the
results of Figures \ref{fig:lstarngals} and \ref{fig:lbcg_ngals}:
\Lbcg\ monotonically increases with \Ntwo, while \Lstarsat\ is
essentially flat for \Ntwo\ $\ge 10$ but correlated with richness for
lower \Ntwo\ systems. We caution that the low richness (\Ntwo\ $< 10$)
systems should not be taken as representative of {\it all} low-mass
systems, as the selection function imposed by the cluster finder
(demanding only a few red galaxies in close proximity) may be
important in this regime.

For \Ntwo\ $\ge 10$, the ratio \Lbcg/\Lstarsat\ as a function of
cluster richness can be described by a power law:
\begin{eqnarray}
\frac{^{0.25}L_{BCG}}{^{0.25}L_*(sat)} = (2.8 \pm 0.2)\, N_{200}^{0.22 \pm 0.05},
\end{eqnarray}
and likewise as a function of mass:
\begin{eqnarray}
\frac{^{0.25}L_{BCG}}{^{0.25}L_*(sat)} = (7 \pm 3) \left(\frac{M_{200}}{M_{14}}\right)^{0.18 \pm 0.02}.
\end{eqnarray}
This relationship is shown in the Figure with a solid line over the range in
\Ntwo\ where the fit was performed and a dashed line to extrapolate the trend
to lower richness.

\section{Summary \& Dicussion}\label{sec:conclusion}
In this study, we have examined the properties of cluster-associated
galaxies, separating BCGs from satellites and focusing on trends in
galaxy color and luminosity as a function of cluster richness,
distance from cluster center, and redshift. We use clusters in the
SDSS \maxbcg\ sample, the largest set of clusters identified to date.
Employing photometric data alone, we apply cross-correlation
background-correction techniques to characterize the
cluster-associated galaxy population, within $5\times$\rtwo\ and
brighter than \Mi\ $< -19$, around \nclust\ systems spanning more than
two decades in mass in the redshift range $0.1 \le z \le 0.3$.

Our principle results are as follows.
\begin{enumerate}

\item The luminosity function of satellites within \rtwo\ as a
    function of cluster mass for systems with mass greater than $3
    \times 10^{13}$\Msun\ shows remarkable uniformity for \Mi\ $< -19$. The
    characteristic satellite luminosity \Lstar\ is only weakly
    dependent on cluster richness.

\item The shape of the luminosity function of satellites brighter than
  \Mi\ $< -19$ does not change with cluster-centric radius. However,
  the color-separated luminosity functions of satellites as a function
  of \rad\ and of \Ntwo\ show that the mix of sub-\Lstar\ red and blue
  galaxies changes dramatically as a function of radius. In contrast,
  the relative number of red and blue galaxies at the bright end is
  roughly constant with radius.

  \item The average color of satellite galaxies is redder near cluster
    centers, but this trend is largely a reflection of the changing
    ratio of red to blue galaxies mentioned above.  This effect is a very
    weak function of cluster mass over the range investigated here.

  \item The fraction of red galaxies increases with cluster mass over
    the full range explored, although only weakly for \Mtwo $\gae
    10^{14}$\Msun.  This fraction decreases with cluster-centric
    distance until $2\times$\rtwo, decreases with luminosity for
    $L<$ \Lstar, and is constant for $r>2\times$\rtwo\ and $L>$
    \Lstar.

  \item The fraction of cluster galaxies within \rtwo\ that are red
    increases by $\sim$5\% during the 0.8 Gyr between redshift
    $z=0.28$ to $z=0.2$; this change is roughly independent of mass
    over the range investigated.

  \item The luminosity of BCGs and the ratios of BCG luminosity to
    total cluster luminosity and to characteristic satellite
    luminosity are all correlated strongly with cluster mass, and we
    have quantified each of these scalings over the mass range $3
    \times 10^{13}$\Msun\ to $9 \times 10^{14}$\Msun.  The BCG
    luminosity has a Gaussian distribution at fixed cluster richness,
    with dispersion $\sigma_{logL} \sim 0.17$ for clusters with \Mtwo\
    $ > 10^{14}$\Msun.

\end{enumerate}

While these results are in general agreement with previous
observational work, due to the volume probed we are able to
investigate a wider mass range, extending the statistics to higher
mass than previous samples.  We are also able to split the sample into
finer bins for several variables than has previously been possible.

The \maxbcg\ selection function, which is well-understood for most of the richness range used here, is less well quantified for clusters
with \Ntwo\ $< 10$. This low richness set of clusters may be less
complete and pure, which could result in the significant difference in
LF shape as compared to higher \Ntwo\ systems, or contribute to the
drop-off in \fred\ for low richness systems. However, not all cluster
galaxy properties change significantly at \Ntwo\ $= 10$ (\eg,
\Lbcg/\Ltwo) and there is no break at a particular \Ntwo\ in either
the mass--observable relationship \citep{Johnston07b} or mass-to-light
ratios \citep{Sheldon07b} as measured by lensing. Interestingly, it is
the quantities that most closely trace the total mass of the systems
(\ie, \Lbcg, \Ltwo and the lensing signal) that are smoothly scaling
over the full richness range, while quantities that are related to the
mix of galaxies within clusters (\ie, the LF and \fred) are the ones
that change more dramatically. We hypothesize that, at these low
richnesses, \maxbcg\ is finding legitimate low-mass systems, but that
the selection priors demanding the close proximity of only a few red
galaxies result in finding only systems with the observed mix of
galaxies and not necessarily {\em all} systems of this low
mass. Further investigation using lower mass threshold mock catalogs
is needed to understand in detail the selection of low-richness
systems.

Our results can shed light both on the processes that build up the
galaxy population in clusters and distinguish central galaxies from
satellites, as well as the processes that are responsible for the
galaxy transformation from blue to red.  With respect to the former,
the results presented here fit well within the basic picture of galaxy
formation in CDM: that galaxy properties are likely linked to the
formation history of a cluster's dark matter halo and its
substructures.  Although detailed comparisons are beyond the scope of
this work, our results qualitatively match both HOD constraints from
clustering statistics as well as models based on matching the
abundance of halos and subhalos to galaxies.  The HOD framework
provides a way to examine the galaxy population in both the observed
Universe and in models of galaxy formation and evolution. Without
needing to identify specific groups or clusters in the data, halo
model interpretations of the statistics of luminosity-dependent galaxy
clustering result in specific predictions for trends of both central
and satellite galaxies as a function of halo mass that are in
reasonable agreement with the findings presented here, especially for
the relationship between central and satellite luminosities
\citep[\eg,][]{Berlind03,Skibba07}.  Models which link the properties
of galaxies directly to their halos and subhalos also agree broadly
with several of the results presented here \citep[\eg,][]{
Conroy06, VO07}.  Detailed predictions for several cluster statistics from
such a model will be presented in \citet{Iro07}.  Our results
on scatter in the BCG luminosity at fixed cluster richness very likely
provide an upper limit on scatter in central galaxy luminosity at
fixed mass, as the scatter between halo mass and cluster richness
should act to increase this scatter.  The fact that this scatter is
already fairly small provides further support for the tight coupling
between halo mass and galaxy luminosity that is the basis of these
models.

In addition to processes that cause physical changes to
satellites resulting from interactions with the cluster gas, cluster
potential, or other satellites, presumably some of the satellites are
lost due to being accreted onto the BCG. In detail, this process
likely results in the stellar component of the disrupted galaxy
joining both the BCG and ICL \citep{Conroy07ICL}, but nonetheless
should result in a BCG population closely linked to both halo mass and
satellite population, as is observed. Indeed, \Lbcg/\Ltwo\ and
\Lbcg/\Lstarsat\ must be intimately related to the processes
responsible for BCG growth. That BCGs get brighter as a function of
cluster mass faster than do typical satellies may be further evidence
that BCGs are different (in their merger history) than typical
satellites.

Understanding the timescales and mechanisms for galaxies to transform
from star-forming galaxies onto the red sequence is one of the primary
current challenges for galaxy formation theories.  There are several
processes that can operate within clusters to shape the population of
the cluster galaxies, such as ram pressure stripping, harassment and
strangulation, that directly influence the galaxies' gas content and
thus their subsequent star formation \citep[for a recent review,
see][]{DeLucia06}. Clearly many of the processes responsible for this
transformation are related either to the mass of the host halo or how
long the galaxies have been satellites.  The relative strengths of
various effects are still rather uncertain, however.  What fraction of
galaxies become red while they are central galaxies?  Do the processes
happen only for satellites in a certain mass range, or do they happen
equally for all satellite galaxies?  On what timescales do these
processes operate?  Our measurements of how the red fraction scales
with cluster mass, radius and redshift will be instrumental in
answering these questions.

We address a few of these issues here.  Ram pressure stripping
predicts that \fred\ will be inversely correlated with cluster-centric
distance, and larger for both brighter galaxies and more massive
halos. However, our results indicate that \fred\ is essentially
independent of galaxy luminosity at fixed \Ntwo\ for galaxies brighter
than \Lstar\ and that the radial trend in \fred\ is not any more
pronounced in high \Ntwo\ systems. These results indicate that ram
pressure stripping cannot be the dominant process at work to transform
the galaxy population.  The harassment scenario predicts that \fred\
will be anticorrelated with cluster mass, as is observed; however, if
this mechanism were dominant, then at fixed cluster mass \fred\ would
be expected to be larger for less luminous galaxies
\citep{Weinmann06a}, contradicting the observed trends. Strangulation,
where star formation in infalling galaxies is halted because no
further gas accretion is allowed, makes several predictions that are
in good agreement with our observations.  For example, the model
presented by \citet{Diaferio01} predicts: that the mean satellite
color gets bluer as a function of cluster-centric distance until
reaching a plateau at the field value around 2--3\rtwo; that the mean
satellite color depends on halo mass only for $M \lae 5 \times
10^{13}$\Msun; and that the incidence of blue galaxies in clusters
increases at higher redshift. However, this model also predicts that
both bulge- and disk-dominated satellies will get redder toward the
central regions of clusters, a trend that we do not see in our
red/blue sample split (or course, our color-separated subsamples are
not directly comparable to their morphology-separated samples).  Note
that the predictions of such a model will in detail depend on its
implementation.  Our observation that the satellite red fraction
changes in the same way with redshift regardless of cluster richness
is significant evidence that the timescale of the physics responsible
for quenching is the same in systems over the full mass range
examined. The difference in cosmic time between the median redshifts
of our two samples is approximately the dynamical timescale, over
which we observe a $\sim$5\% change in \fred, and this observation
should allow for useful constraints in studies on the details of
quenching mechanisms.

A consensus is emerging from a variety of modeling efforts that star
formation proceeds most efficiently in a mass range around $L_*$, and is
less efficient in more massive halos and for satellite galaxies.
Simple models based on this type of assumption can produce many of the
rough trends that we have seen here; for example, that the mean color
of galaxies will not change significantly as a function of halo mass
for the range of masses we investigate here \citep[\eg,][]{Diaferio01}, and
that the galaxy type fraction should be a weak function of halo mass
for $M$ \gae $10^{13}$\Msun\ \citep{Berlind03,Zheng05,Cooray05}.
Furthermore, the general concept of galaxies falling in, quenching,
and fading matches well with the observed radial trends of LF and red
fraction.  However, most of the detailed semi-analytic modeling
efforts have had some trouble matching in detail observations of the
color distribution of galaxies and how it changes with environment
\citep[see \eg,][]{Coil07}.  Unresolved issues include the rates at
which star formation is triggered or shut off, and accordingly red
galaxies tend to be overproduced.  To understand in detail the
physical mechanisms responsible for the quenching of star formation as
clusters are assembled, further work is needed to accurately reproduce
the observed trends in the cluster galaxy population.  The present
findings set a local-universe target for modeling results, and provide
some guidance for the relative importance of some of the germane
effects.

A substantial effort over the next decade will be devoted to
large-scale, multi-band photometric surveys, including DES, Pan-STARRS,
and LSST. Although the primary science driver of many of these projects is
to investigate the nature of dark energy, the resulting data are
likely to also provide strong constraints on the processes of galaxy
evolution. The results presented here provide a low-redshift baseline
against which current and future high-redshift samples may be
compared.  From a technical standpoint, these data are informative to
next-generation cluster-finding techniques, and are useful input for
creating the mock catalogs necessary for interpreting cluster surveys.
Furthermore, the techniques presented here, which use photometric data
alone, are directly applicable to these upcoming imaging surveys, and
will thus enable detailed studies of the galaxy population at
significantly higher redshifts without extensive spectroscopy.

\acknowledgments This project was supported by the Kavli Institute for
Cosmological Physics (KICP) at the University of Chicago. SMH
appreciates the hospitality of Anja von der Linden while part of this
work was being completed. RHW was supported in part by the
U.S. Department of Energy under contract number DE-AC02-76SF00515 and
by a Terman Fellowship from Stanford University.  We thank Rebecca
Bernstein for discussion about detection thresholds, Brian Gerke and
Charlie Conroy for helpful comments on the manuscript, and Andreas
Berlind, Frank van den Bosch, and Tim McKay for useful conversations.

Funding for the SDSS and SDSS-II has been provided by the Alfred
P. Sloan Foundation, the Participating Institutions, the National
Science Foundation, the U.S. Department of Energy, the National
Aeronautics and Space Administration, the Japanese Monbukagakusho, the
Max Planck Society, and the Higher Education Funding Council for
England. The SDSS Web Site is http://www.sdss.org/.
The SDSS is managed by the Astrophysical Research Consortium for the
Participating Institutions. The Participating Institutions are the
American Museum of Natural History, Astrophysical Institute Potsdam,
University of Basel, University of Cambridge, Case Western Reserve
University, University of Chicago, Drexel University, Fermilab, the
Institute for Advanced Study, the Japan Participation Group, Johns
Hopkins University, the Joint Institute for Nuclear Astrophysics, the
Kavli Institute for Particle Astrophysics and Cosmology, the Korean
Scientist Group, the Chinese Academy of Sciences (LAMOST), Los Alamos
National Laboratory, the Max-Planck-Institute for Astronomy (MPIA),
the Max-Planck-Institute for Astrophysics (MPA), New Mexico State
University, Ohio State University, University of Pittsburgh,
University of Portsmouth, Princeton University, the United States
Naval Observatory, and the University of Washington.  This work made
extensive use of the NASA Astrophysics Data System and the {\tt
  arXiv.org} preprint server.


\end{document}